\documentclass[prd,preprint,eqsecnum,nofootinbib,amsmath,amssymb,
               tightenlines,dvips]{revtex4}
\usepackage{graphicx}
\usepackage{bm}

\newif\iffrivolous \frivolousfalse	
\def\brem{bremsstrahlung}
\def\lra{\leftrightarrow}
\def\degen{\nu}
\def\dpslash{\frac{d^3\p}{(2\pi)^3} \>}

\def\Re{{\rm Re}}

\def\PiT{\Pi_{\rm T}}

\newcommand\Eq[1]{Eq.~(\ref{#1})}
\def\onetwo{{1\lra2}}
\def\twotwo{{2\lra2}}
\def\pressure{{\cal P}}
\def\mg{m_{{\rm eff,g}}}
\def\mf{m_{{\rm eff,f}}}
\def\meffs{m_{{\rm eff},s}}
\def\Qmax{Q_{\rm max}}
\def\ul{\underline}

\def\h{{\bm h}}
\def\j{{\bm j}}
\def\x{{\bm x}}
\def\p{{\bm p}}
\def\q{{\bm q}}
\def\k{{\bm k}}
\def\v{{\bm v}}
\def\u{{\bm u}}
\def\E{{\bm E}}

\def\pxk{\h}

\def\A{{\cal A}}

\def\F{{\bm F}}
\def\FF{\tilde F}

\def\naBla{{\bm \nabla}}
\def\ij{{i \cdots j}}

\def\ca{C_{\rm A}}
\def\cf{C_{\rm F}}
\def\da{d_{\rm A}}

\def\df{d_{\rm F}}
\def\nf{N_{\rm f}\,}
\def\nc{N_{\rm c}\,}

\def\mD{m_{\rm D}}
\def\mQ{m_{\rm F}}

\def\trans{\top}

\def\bSigma{{\bm \Sigma}}

\def\CLL{{\cal C}_{\rm LL}}

\def\Fext{\F_{\rm ext}}

\def\grad{\mbox{\boldmath$\nabla$}}
\def\half{{\textstyle{\frac 12}}}
\def\quarter{{\textstyle{\frac 14}}}

\def\alphas{\alpha_{\rm s}}
\def\alphaEM{\alpha_{\rm EM}}

\def\C{{\cal C}}
\def\S{{\cal S}}

\def\la{\label}

\def\slashchar#1{\setbox0=\hbox{$#1$}           
   \dimen0=\wd0                                 
   \setbox1=\hbox{/} \dimen1=\wd1               
   \ifdim\dimen0>\dimen1                        
      \rlap{\hbox to \dimen0{\hfil/\hfil}}      
      #1                                        
   \else                                        
      \rlap{\hbox to \dimen1{\hfil$#1$\hfil}}   
      /                                         
   \fi}                                         %

\def\lsim{\mbox{~{\raisebox{0.4ex}{$<$}}\hspace{-1.1em}
	{\raisebox{-0.6ex}{$\sim$}}~}}

\def\Dq{D_{\rm q}}
\def\Dlangle{\left\langle\!\!\left\langle}
\def\Drangle{\right\rangle\!\!\right\rangle}
\def\qt{\q^\perp}
\def\QLPM{{\cal Q}}
\def\KLPM{{\cal K}}
\def\mom{r}                        
\def\dCbar{\overline {\delta \C}}

\advance\parskip 2pt

\begin {document}


\preprint {UW/PT 03--01}

\title
    {
    Transport coefficients in high temperature gauge theories:\\
    (II) Beyond leading log
    }

\author {Peter Arnold}
\affiliation
    {%
    Department of Physics,
    University of Virginia,
    Charlottesville, VA 22901
    }%
\author{Guy D. Moore}
\affiliation
    {%
    Department of Physics,
    McGill University, 
    3600 University St.,
    Montr\'{e}al, QC H3A 2T8, Canada
    }%
\author{Laurence G. Yaffe}
\affiliation
    {%
    Department of Physics,
    University of Washington,
    Seattle, Washington 98195
    }%

\date {February 2003}

\begin {abstract}%
    {%
    Results are presented of a full leading-order evaluation of
    the shear viscosity, flavor diffusion constants,
    and electrical conductivity
    in high temperature QCD and QED.
    The presence of Coulomb logarithms associated with gauge interactions
    imply that the leading-order results for transport coefficients
    may themselves be expanded in an infinite series in powers of
    $1/\log(1/g)$;
    the utility of this expansion is also examined.
    A next-to-leading-log approximation is found to approximate the
    full leading-order result quite well as long as the Debye mass is
    less than the temperature.
    }%
\end {abstract}

\maketitle
\thispagestyle {empty}

\section {Introduction}
\la{sec:intro}

    Transport coefficients, such as viscosities, diffusivities, or
electric conductivity, characterize the dynamics of long wavelength,
low frequency fluctuations in a medium.
In a weakly coupled quantum field theory,
transport coefficients should, in principle, be calculable 
purely theoretically.
The values of transport coefficients are of interest
in cosmological applications such as electroweak baryogenesis
\cite{baryo1,baryo2}
and the origin of primordial magnetic fields \cite{mag_fields},
as well as for hydrodynamic models of heavy ion collisions
(see, for example, Refs.\ 
\cite {teaney,heavy-ion1,heavy-ion2,heavy-ion3,heavy-ion4a,heavy-ion4b,heavy-ion4c},
and references therein).
{}From a purely theoretical perspective, the evaluation of transport
coefficients also provides an excellent test of our understanding of
dynamic processes in thermal field theory.

In a previous paper \cite{AMY1}, we performed
leading-log calculations of the shear viscosity,
electrical conductivity, and flavor diffusion constants
in weakly coupled, high temperature gauge theories%
\footnote
    {%
    The leading-log result for shear viscosity in pure Yang-Mills
    theory was first evaluated by Heiselberg \cite {Heiselberg}
    and closely approximated by earlier work of Baym {\em et al.}\/
    \cite{BMPRa}.
    A variety of incorrect leading-log calculations of transport
    coefficients are also present in the literature;
    see Ref.~\cite{AMY1} for a discussion of this earlier work.
    }%
---that is, neglecting relative corrections suppressed only by
powers of the inverse logarithm of the gauge coupling, $1/\log(1/g)$.
Leading-log calculations may be regarded as improvements over
phenomenological estimates based on relaxation time approximations
(see, for example, Refs.
\cite {HosoyaKajantie,Hosoya_and_co,relax1,relax2,relax3,relax4}),
but no leading-log calculation can be trusted to provide
even a factor of two estimate in any real application,
because the logarithm of the inverse gauge coupling,
even for electromagnetism,
is never all that large.

At a minimum, one would like to know several terms in the
expansion in inverse powers of $\log(1/g)$ in order to assess
the utility of this asymptotic series.
Even better would be a full {\em leading-order} calculation of
transport coefficients, by which we mean an evaluation which
correctly includes all effects suppressed by powers of
$1/\log(1/g)$ and only neglects relative corrections suppressed
by powers of $g$.
This is a feasible goal.
Transport coefficients are dominantly sensitive%
\footnote
    {
    This is true of the transport coefficients under consideration
    in this paper.
    It should be noted that bulk viscosity requires a separate
    treatment because it vanishes identically in a scale invariant theory,
    and so requires the correct incorporation of
    effects which are irrelevant for other transport coefficients.
    For a discussion of bulk viscosity in the case of scalar field theory,
    see Refs.~\cite{Jeon,JeonYaffe}.
    }
to the dynamics of excitations ({\em i.e.}, quarks and gluons)
with typical momenta of order of the temperature $T$.
And, as we recently discussed in Ref.~\cite {AMY5},
it is possible to formulate an effective kinetic
theory which correctly describes the leading-order dynamics
of such excitations.
The purpose of this paper is to perform a full leading-order
evaluation of shear viscosity, electrical conductivity,
and flavor diffusion constants, and also to examine the utility
of the asymptotic expansion in inverse powers of $\log(1/g)$.
We believe this work represents the first reliable analysis
of transport coefficients beyond a leading-log approximation
in high temperature gauge theories.

We begin in Section~\ref{sec:review}
by summarizing the relevant definitions,
reviewing the form of the effective kinetic theory,
and sketching how the actual calculation of transport coefficients
reduces to a linear integral equation which may be accurately
solved using a variational approach.
The evaluation of matrix elements of the linearized collision operator
is discussed in Appendices~\ref {app:22} and \ref {app:12}.
Our results for the leading-order fermion diffusion constant
and shear viscosity in QCD
are presented in Section~\ref{sec:leading-order},
and the expansion in powers of $1/\log(1/g)$
is discussed in Section~\ref{sec:log-expansion}.
We include here next-to-leading-log results for both Abelian
and non-Abelian theories.
We find that a next-to-leading-log approximation reproduces
the full leading-order results substantially better than one
might have expected.
Section~\ref{sec:elec-cond} presents both leading-order,
and next-to-leading-log results for
the electrical conductivity in QED (or in quark plus lepton) plasmas.
A short conclusion summarizes our results and comments briefly on
the relation between our effective kinetic theory treatment and
purely diagrammatic approaches.
Several more technical appendices follow.
The first two present details of the required $2\lra2$ particle
effective scattering rates, and the $1\lra2$ particle effective
splitting rates which characterize nearly collinear \brem\ and 
pair production processes in the fluctuating
soft thermal gauge field background.
A final appendix contains a proof that the expansion of leading-order
transport coefficients in inverse powers of $\log(1/g)$
is only asymptotic, not convergent,
and gives a non-rigorous argument suggesting that this expansion
is not Borel summable.

\section{Ingredients}
\label{sec:review}

Throughout this work, we assume that the gauge coupling
(defined at the scale of the temperature) is weak, $g(T) \ll 1$.
For QCD, this means that the temperature $T$ is asymptotically large
compared to $\Lambda_{\rm QCD}$.
We assume that there are no particle masses close to $T$.
Zero-temperature masses must either be negligible compared to $gT$
(so thermal self-energies are large compared to the mass),
or large compared to $T$
(so the particle decouples and may be ignored completely).
In particular this means we will not consider temperatures
just below thermal phase transitions (or crossovers).
We also assume that any chemical potentials for conserved numbers
are small compared to $T$.

\subsection{Definitions}

Transport coefficients characterize a system's response to weak, slowly
varying inhomogeneities or external forces.  
If a weak and spatially uniform electric field $\E$ is applied to a plasma,
the resulting induced electric current is
\begin{equation}
    \langle {\j}^{\rm EM} \rangle = \sigma \, \langle \E \rangle \,,
\la {eq:jiEM}
\end {equation}
where the electric conductivity $\sigma$ is a function of the temperature $T$,
the electric charge $e$,
and the particle content of the theory.
This constitutive relation is satisfied up to corrections involving gradients 
of the electric field and higher powers of the field strength;
strictly speaking the conductivity is defined by the above relation
in the limit of vanishing frequency and wavenumber of an
infinitesimal applied electric field.

In a theory with a conserved global charge (such as baryon or lepton number),
the associated charge density $n \equiv j^0$ and current density $\j$
will satisfy a diffusion equation,
\begin {equation}
    \langle {\j} \rangle
    = -D \> \naBla \langle n \rangle \,,
\la {eq:jia}
\end {equation}
in the local rest frame of the medium.%
\footnote
    {%
    We work throughout in the local rest frame of some arbitrary point $x$
    in the system.
    This frame is defined by the Landau-Lifshitz convention of
    vanishing momentum density, $\langle T^{0i}(x) \rangle=0$.
    }  
The coefficient $D$ is called the diffusion constant.

When one or more diffusing species of excitations carry electric charge,
the electric conductivity is connected to the diffusion constants for
various species by an Einstein relation
(see Ref.~\cite {AMY1} for more discussion).
If the net number of each species of charge carriers is conserved, then
\begin {equation}
    \sigma = \sum_a \, e_a^2 \, D_a \, \frac {\partial n_a}{\partial \mu_a} \,,
\label {eq:einstein}
\end {equation}
where the sum runs over the different species or flavors of excitations
with $e_a$ and $D_a$ the corresponding electric charge and diffusion constant,
respectively.%
\footnote
    {%
    For a single massless Dirac fermion at vanishing chemical potential,
    the susceptibility $\partial n_a/\partial \mu_a = \frac 13 \, T^2$
    (at leading order in the gauge coupling).
    For quarks, this should be multiplied by 3 (or the number of colors).
    However, the electrical conductivity is dominated by the transport
    of charged leptons, which diffuse more readily than quarks since
    they only interact electromagnetically.
}

If the flow velocity of the plasma is not uniform, then the
stress tensor (which defines the flux of momentum density)
will depart from its perfect fluid form.
In the local fluid rest frame at a point $x$,
the stress tensor, to lowest nontrivial order in the velocity gradient,
will have the form
\begin {equation}
    \langle T_{ij}(x) \rangle
    =
    \delta_{ij} \, \langle {\pressure} \rangle
    - \eta \left[ \nabla_i \, u_j + \nabla_j \, u_i
		- {\textstyle \frac 23} \, \delta_{ij} \, \nabla^l \, u_l
	    \right]
    - \zeta \, \delta_{ij} \, \nabla^l \, u_l \,,
\la {eq:Tij}
\end {equation}
where ${\pressure}$ is the equilibrium pressure associated with
the energy density $\langle T_{00}(x) \rangle$,
and the coefficients $\eta$ and $\zeta$ are known as
the shear and bulk viscosities, respectively.
The flow velocity $\u$ equals the momentum density divided by the
enthalpy density (which is the sum of energy density and pressure).
We will only be concerned with the shear viscosity in this paper;
the bulk viscosity requires a more complicated analysis,
which to date has only been performed for a scalar field theory
\cite{Jeon,JeonYaffe}.  

\subsection{Effective kinetic theory}

In a weakly coupled plasma, the energy, momentum, electric charge,
and other global conserved charges, are predominantly carried by
excitations with momenta $p \gg gT$.
Such momenta will be referred to as ``hard.''
Weak coupling implies that these excitations are
long-lived (and hence well-defined) quasiparticles.
Thermal corrections to the dispersion relations of excitations
are order $gT$ in size [{\em i.e.}, $(p^0)^2 = \p^2 + O(g^2T^2)$].
Hence, hard excitations are ultrarelativistic.

The parametrically large separation between quasiparticle
lifetimes and the screening length of order $1/(gT)$
ensures that the dynamics of these excitations
may be reproduced by an effective kinetic theory.%
\footnote
    {
    In particular, the contributions of hard degrees of freedom
    to the flux of energy, momentum, or other global conserved charges
    are captured by the two-point Wigner functions for the hard degrees
    of freedom,
    up to sub-leading corrections.
    As discussed in, for instance,
    Refs.~\cite{Jeon,JeonYaffe,Blaizot&Iancu,CalzettaHu1,CalzettaHu2,A&Y},
    kinetic theory correctly describes the time evolution of the two-point
    function to within $O(g)$ or smaller corrections.
    }
In other words, one may describe the instantaneous state of the
system by a set of phase space distribution functions $f^a(\x,\p,t)$.
The index $a$ will label the relevant species of excitations
({\em i.e.}, gluon, up-quark, up-antiquark, down-quark, down-antiquark,
{\em etc.}, for QCD).
Formally, the distribution functions represent spatially smeared
and Wigner transformed non-equilibrium two-point correlation functions.
The evolution of
phase space distribution functions is governed by effective
Boltzmann equations of the form
\begin {equation}
    \left[
	\frac{\partial}{\partial t}
	+
	\v_\p \cdot \frac{\partial}{\partial \x}
	+
	\Fext^a \cdot \frac{\partial}{\partial \p}
    \right]
    f^a(\p,\x,t)
    =
    -C_a[f] \, ,
\la {eq:Boltz}
\end {equation}
where $\v_\p \equiv \partial p^0/\partial \p$ is the velocity of an
excitation with momentum $\p$,
$\Fext^a(\x,t)$ is an external force acting on excitations of type $a$
(due, for example, to an applied electric field),
and $C_a[f]$ is the collision integral
which characterizes the net rate at which species $a$ excitations with momentum
$\p$ are lost (or created) due to scattering processes involving
other excitations in the plasma.
For ultrarelativistic hard excitations, the velocity is a unit vector,
$\v_\p = \hat \p$, up to irrelevant subleading corrections.

As discussed at length in Ref.~\cite{AMY5}
and proposed earlier by Baier, Mueller, Schiff, and Son \cite{BMSS},%
\footnote
    {%
    See also Refs.~\cite{Gelis1,AMY2,AMY3,AMY4,AGMZ} for discussion of
    the closely related processes which contribute to the leading-order
    photon emission rate.
    }
the collision term needed to reproduce leading-order quasiparticle
dynamics must contain both $2\lra2$ particle scattering terms
and effective $1\lra2$ particle splitting and joining terms.
The latter represent nearly-collinear \brem\ and pair production/annihilation
processes which take place in the presence of soft ($gT$ scale)
thermal fluctuations in the background gauge field.
Hence, the collision terms take the form
\begin {equation}
    C_a[f] = C_a^\twotwo[f] + C_a^\onetwo[f] \,,
\end {equation}
where
\begin {eqnarray}
    C_a^\twotwo[f](\p)
    &=&
    \frac {1}{4|\p|\nu_a} \sum_{bcd}
    \int_{\k\p'\k'}
	\left| {\cal M}^{ab}_{cd}(\p,\k;\p',\k') \right|^2 \>
	(2\pi)^4 \, \delta^{(4)}(P + K - P' - K')
\nonumber\\ && \hspace {0.9in} {} \times
    \Bigl\{
	f^a(\p) \, f^b(\k) \, [1{\pm}f^c(\p')] \, [1{\pm}f^d(\k')]
\nonumber\\ && \hspace {1.0in} {}
	-
	f^c(\p') \, f^d(\k') \, [1{\pm}f^a(\p)] \, [1{\pm}f^b(\k)]
    \Bigr\} \,,
\la {eq:collision22}
\\
\noalign{\hbox{and}}
    C_a^\onetwo[f](\p)
	&=& \frac{(2\pi)^3}{2|\p|^2\nu_a} \sum_{bc}
	\int_0^\infty dp'\> dk' \;
	\delta (|\p| - p' - k' ) \;
	\gamma^{a}_{bc}(\p;p'\hat\p,k'\hat\p)
\nonumber\\ && \hspace{3em} {} \times
    \Bigl\{
	f^a(\p) \, [1{\pm}f^b(p' \hat\p)] \, [1{\pm}f^c(k' \hat\p)]
	-
	f^b(p' \hat\p) f^c(k' \hat\p) \, [1{\pm}f^a(\p)]
    \Bigr\}
\nonumber \\ &+&
	\frac {(2\pi)^3}{|\p|^2\nu_a} \sum_{bc}
	\int_0^\infty dk \> dp' \; 
	\delta (|\p| + k - p' ) \;
	\gamma_{ab}^{c}(p'\hat \p;\p,k \, \hat\p)
\nonumber\\ && \hspace{3em} {} \times
    \Bigl\{
	f^a(\p) \, f^b(k \hat\p) [1{\pm}f^c(p' \hat\p)]
	-
	f^c(p' \hat\p) \, [1{\pm}f^a(\p)][1{\pm}f^b(k \hat\p)]
    \Bigr\}.\quad
\la {eq:collision12}
\end {eqnarray}
Here, $P=(p^0,\p)$, $K=(k^0,\k)$, {\em etc}.\ denote null four-vectors
(so that $p^0 \equiv |\p|$, {\em etc.}),
$\nu_a$ is the number of spin times color states for species $a$
({\em i.e.}, 6 for each quark or antiquark and 16 for gluons),
and $\int_\p \equiv \int d^3\p / [2|\p| \, (2\pi)^3]$ denotes
Lorentz-invariant momentum integration.
As usual, upper signs refer to bosons and lower signs to fermions.
${\cal M}^{ab}_{cd}(\p,\k;\p',\k')$
is the effective two body scattering amplitude
for the process $ab \lra cd$, defined with a relativistic normalization
for single particle states;
its square $\left| {\cal M}^{ab}_{cd} \right|^2$ is implicitly
understood to be summed (not averaged) over the spins and colors
of all four excitations.
Similarly,
$\gamma^{a}_{cd}(\p;p' \hat\p,k' \hat\p)$
is the differential rate for an $a \to bc$
effective splitting process (or its time reverse),
integrated over the parametrically small transverse momenta
of the participants.
This rate is likewise understood to be
summed over the spins and colors of all three participants.
The prefactors of $1/(4|\p|)$ in the $\twotwo$ terms
are a combination of the $1/(2|\p|)$ 
from the relativistic normalization of scattering amplitudes
together with a symmetry factor of 1/2 which 
corrects for double counting of final or initial states
(specifically $\p',c$ interchanged with $\k',d$).

The $2\lra2$ matrix elements
$| {\cal M}^{ab}_{cd}|^2$
are the usual lowest order vacuum amplitudes, except that
thermal self-energies must be included on internal gauge boson
or fermion lines in $t$ and $u$ channel exchange processes \cite{AMY5}.
These thermal self-energies are evaluated in the hard thermal loop (HTL)
approximation, which is adequate for a leading-order analysis.
Explicit expressions may be found in Ref.~\cite {AMY5}
and also appear in Appendix \ref {app:22}, which discusses
the evaluation of the $\twotwo$ collision terms.

The $1\lra2$ splitting rates
$\gamma^{a}_{cd}$
encapsulate the effect of near-collinear processes
in which any number of soft scatterings with other excitations in the
plasma occur during the emission.
The importance of including such multiple soft scatterings is known as
the Landau-Pomeranchuk-Migdal (LPM) effect.
It leads to a two-dimensional linear integral equation which must be
solved, for each value of $|\p|$ and $k$, to determine the
$1\lra2$ transition rate $\gamma^a_{bc}$.
This is discussed further in Ref.~\cite {AMY5}
and in Appendix \ref {app:12}, which describes the evaluation
of these $\onetwo$ splitting rates.

\subsection {Linearization}

Given an effective kinetic theory, transport coefficients
are calculated by linearizing the Boltzmann equation
about local equilibrium.
The required procedure is discussed in some detail in our
earlier paper \cite{AMY1}.
In brief,
one writes each distribution function as a slowly varying local
equilibrium part $f^a_0(\x,\p)$
plus a small departure from (local) equilibrium
$f^a_1(\x,\p)$.
For electric conductivity, $f^a_0$ may equal a homogeneous
equilibrium distribution.
For diffusion, $f^a_0$ is a local equilibrium distribution in which
the chemical potential coupled to the conserved charge of interest
varies slowly in space.
For shear viscosity, $f^a_0$ is a local equilibrium distribution in which
the flow velocity has a non-zero shear.
In each case,
the leading contribution on the left side of the Boltzmann equation
(\ref{eq:Boltz})
comes from either the convective derivative or external force term
acting on $f^a_0$.
Because the collision terms vanish identically for any local
equilibrium distribution,
the leading contribution on the right side of the equation
is linear in the departure $f^a_1$ from equilibrium.
One solves the resulting linearized Boltzmann equation for $f^a_1$,
uses the result to evaluate the stress tensor or current density of interest,
and then reads off the appropriate transport coefficient.

The convective derivative acting on $f^a_0$ is
proportional to the relevant driving term, which we will denote by
\begin {equation}
    X_\ij
    \equiv
    \begin {cases}
	-E_i \,, & \mbox{(conductivity),} \cr
	\;\nabla_i \, \mu_\alpha \,, & \mbox{(diffusion),} \cr
	\frac{1}{\sqrt{6}} \left( \nabla_i u_j + \nabla_j u_i
	- \frac{2}{3} \, \delta_{ij} \nabla \cdot \u \right) \,, 
	& \mbox{(shear viscosity).} \cr
    \end {cases}
\label {eq:Tdrive}
\end {equation}
This tensor has $\ell=1$ angular dependence for conductivity and diffusion,
and $\ell=2$ angular dependence for shear viscosity.
The departure from equilibrium which solves the linearized
Boltzmann equation must be proportional
to $X_\ij$.
It is convenient to express the departure in the local rest frame
(where $f_0^a$ is isotropic) as
\begin{equation}
    f^a_1(\p) = \beta^2 f^a_0(p) [1{\pm}f^a_0(p)] \,
	X_\ij \, \chi_\ij^a(\p) \,,
\label {eq:f1}
\end{equation}
which defines $\chi_\ij^a(\p)$.
The resulting linearized Boltzmann equation for the functions
$\chi_\ij^a(\p)$ may be written compactly in the form
\begin{equation}
    \S^a_\ij(\p) = (\C \chi_\ij)^a ( \p) \, ,
\label{eq:Boltz1}
\end{equation}
where $\C$ is a linearized collision operator defined below.
The source term is
\begin{equation}
\S^a_\ij(\p) \equiv - T q^a f_0^a(p)[1{\pm}f_0^a(p)] \, I_\ij(\hat\p) .
\end{equation}
Here $q^a$ denotes the relevant conserved charge
carried by species $a$ associated with
the transport coefficient of interest,
\begin{equation}
q^a \equiv \begin {cases}
	e^a_{\rm EM} \, , & \mbox{(conductivity),} \cr
	q^a_\alpha \, , & \mbox{(diffusion),} \cr
	|\p| \, , & \mbox{(shear viscosity),} \cr
	\end {cases}
\label{eq:qcases}
\end{equation}
where in the case of diffusion, $\alpha$ is simply a label for the flavor
symmetry of interest ({\em e.g.}\ quark number or lepton number).
$I_\ij$ is the unique $\ell=1$ or $\ell=2$ rotationally covariant
tensor depending only on $\hat\p$,
\begin {equation}
    I_\ij(\hat \p)
    \equiv
    \begin {cases}
	\;\hat p_i \,, & \ell = 1 \mbox{ (conductivity/diffusion),} \cr
	\sqrt {\frac 32} \, (\hat p_i \hat p_j - {\frac 13} \delta_{ij}) \,, 
	& \ell = 2 \mbox{ (shear viscosity).} \cr
    \end {cases}
\label {eq:Iij}
\end {equation}
The normalization on $I_\ij$ was chosen so that
\begin {equation}
   I_\ij(\hat\p) \, I_\ij(\hat\p) = 1
\label {eq:Inorm}
\end {equation}
(which is the reason for the peculiar normalization of $X_{ij}$ in the shear
viscosity case).  More generally,
\begin {equation}
   I_\ij(\hat \p) \, I_\ij(\hat \k) = P_\ell^{}(\hat\p\cdot\hat\k) ,
\label {eq:II}
\end {equation}
where $P_\ell(x)$ is the $\ell$'th Legendre polynomial.
Rotational invariance of the collision operator (in the local rest
frame) implies that
\begin{equation}
\chi^a_\ij(\p) = I_\ij(\hat\p) \> \chi^a(p) \,
\end{equation}
for some scalar function $\chi^a(p)$.
Here and throughout, $p \equiv |\p|$.
Solving the linearized Boltzmann equation means determining
$\chi^a(p)$.

Let $N_s$ be the number of relevant species, and
define an inner product on $N_s$-component functions of momenta,
\begin {equation}
\label{eq:inner_product}
    \Big( f,g \Big) \equiv
    \beta^3 \sum_a \, \degen_a \! \int\dpslash \, f^a(\p) \, g^a(\p) \,.
\end {equation}
One may show that the linearized collision operator $\cal C$ is symmetric
with respect to this inner product.
It is a positive semi-definite operator, which is strictly positive
definite in the $\ell {=} 1$ and $\ell{=}2$ channels relevant
for diffusion or shear viscosity.
Consequently,
the linearized Boltzmann equation (\ref{eq:Boltz1}) is precisely
the condition for maximizing the functional
\begin {equation}
    Q[\chi]
    \equiv
    \Big( \chi_\ij, \S_\ij\Big)
    - \half \, \Big( \chi_\ij, \C \, \chi_\ij \Big) .
\la {eq:Q1}
\end {equation}
The explicit forms of the source and collision parts of this quadratic
functional are
\begin {align}
   \Big( \chi_\ij, \S_\ij \Big)
    &{}=
    -\beta^2 \sum_a
    \int \dpslash 
	f_0(p) \, [1\pm f_0(p)] \>
	q^a \, \chi^a(p) \,,
\label {eq:Sij}
\\
\noalign {\hbox {and}}
    \Big( \chi_\ij, \C \, \chi_\ij \Big)
    &{}=
    \Big( \chi_\ij, \C^\twotwo \, \chi_\ij \Big) +
    \Big( \chi_\ij, \C^\onetwo \, \chi_\ij \Big) \,,
\la {eq:Q2}
\\
\noalign {\hbox {with}}
    \Big( \chi_\ij, \C^\twotwo \, \chi_\ij \Big)
    &{}\equiv
    \frac {\beta^3}8
    \sum_{abcd}
    \int_{\p\k\p'\k'}
	\bigl| {\cal M}^{ab}_{cd}(\p,\k;\p',\k') \bigr|^2 \>
	(2\pi)^4 \, \delta^{(4)}(P{+}K{-}P'{-}K')
\nonumber\\ & \kern1in {} \times
	f^a_0(p) \, f^b_0(k) \, [1{\pm}f^c_0(p')] \, [1{\pm}f^d_0(k')]
\nonumber\\  & \kern1in {} \times
	\Bigl[
	    \chi^a_\ij(\p) + \chi^b_\ij(\k) -
	    \chi^c_\ij(\p') - \chi^d_\ij(\k')
	\Bigr]^2_{\strut} \,,
\la {eq:C22}
\\
    \Big( \chi_\ij, \C^\onetwo \, \chi_\ij \Big)
    &{}\equiv
       \frac{\beta^3}{2}
       \sum_{abc} 
       4\pi
       \int_0^\infty dp' \> dp \> dk \;
       \gamma^a_{bc}(p';p,k) \;
       \delta(p'-p-k)
\nonumber\\ & \kern 1in {} \times
          f_0^a(p') \, [1 {\pm} f_0^b(p)] \, [1 {\pm} f_0^c(k)] \>
	\Bigl[
	    \chi^a(p') - \chi^b(p) - \chi^c(k)
	\Bigr]^2 \,.
\la {eq:C12}
\end {align}
We have used the normalization condition (\ref{eq:Inorm})
to simplify the expression (\ref {eq:C12}).
[One may also use the identity (\ref{eq:II}) to contract the spatial indices
in the $\twotwo$ piece (\ref {eq:C22}).]
The overall coefficients of $1/8$ and $1/2$ in the collision parts
(\ref{eq:C22}) and (\ref{eq:C12})
are symmetry factors which compensate for multiple counting of the
same physical process.
We have dropped the directional information $\hat\p$ in the arguments of
the collinear splitting functions $\gamma^a_{bc}$ because, in the linearized
theory, these are to be evaluated in the background of the isotropic
distribution $f_0(\p)$ and so do not depend on direction.
See Appendix \ref{app:12} for further details.

Maximizing the functional $Q[\chi]$ is equivalent to computing
an expectation value of the inverse of the linearized collision operator,
\begin {equation}
    \Qmax = \half \Bigl( \S_\ij, \, \C^{-1} \, \S_\ij \Bigr) \,.
\label {eq:Qmax1}
\end {equation}
The value of the extremum directly determines the actual transport
coefficients, 
\begin {subequations}
\label {eq:anyQ}
\begin {eqnarray}
    \sigma
    &=&
    {\textstyle {\frac 23}} \,
    \Qmax \Bigr|_{\ell = 1, \; q = q_{\rm EM}} \,,
\la {eq:sigmaQ}
\\[5pt]
    D_\alpha
    &=&
    {\textstyle {\frac 23}} \,
    \Qmax \Bigr|_{\ell = 1, \; q = q_\alpha}
    \left( \frac{\partial n_\alpha}{\partial\mu_\alpha} \right)^{-1} ,
\la {eq:DQ}
\\[5pt]
    \eta
    &=&
    {\textstyle \frac{2}{15}} \,
    \Qmax \Bigr|_{\ell = 2, \; q = |\p|} \,.
\la {eq:etaQ}
\end {eqnarray}
\end {subequations}
The charge susceptibility appearing in expression (\ref {eq:DQ})
for $D_\alpha$ is
\begin{equation}
    \frac{\partial n_\alpha}{\partial \mu_\alpha}
    = \frac{T^2}{12} \sum_a \degen_a \, \lambda_a \, (q_\alpha^a)^2 \, ,
\end{equation}
where $\lambda_a$ is 1 for fermionic and 2 for bosonic species.
If $n_\alpha$ is the net number density of a single
fermion flavor, then this susceptibility is $\frac 13 \, \nc \, T^2$
for a massless Dirac fermion in the fundamental representation of SU($\nc$).

In numerical computations, we will focus in this paper on transport
coefficients at zero chemical potential, so that the local equilibrium
distributions $f_0^a(p)$ appearing in
the pieces (\ref{eq:Sij}) and (\ref{eq:Q2}) of $Q[\chi]$ are simply
\begin {equation}
   f_0^a(p) = \frac{1}{e^{\beta p} \mp 1} \, .
\end {equation}
We will primarily discuss the quark number diffusion constant $\Dq$ and
the shear viscosity $\eta$ in QCD,
although we will also report results for the electrical conductivity
in a QED plasma (with or without quarks).
Because we treat all relevant quarks
as massless, the diffusion constants for individual flavors of quarks
in QCD are all the same and equal to $\Dq$, which is also
the diffusion constant for baryon number.

\subsection {Variational solution}\label {sec:vary}

To maximize the functional $Q[\chi]$ exactly,
one must work in the infinite dimensional space of
arbitrary functions $\chi(p)$.
However, as with many other variational problems, one can
obtain highly accurate approximate results by performing a
restricted extremization within a well chosen finite dimensional subspace.
We will select a finite set of basis functions,
$\{ \phi^{(m)}(p)\}$, $m = 1 \ldots K$, and only consider functions 
$\chi^a(p)$ which are linear combinations of these basis functions,
\begin{equation}
\chi^a(p) = \sum_{m=1}^K \tilde \chi^a_m \; \phi^{(m)}(p) \, .
\end{equation}
(More precisely, we will choose increasingly large sets of basis functions,
in order to study the convergence of results with the size of the
basis.)  Restricted to this subspace,
the source and collision parts of the functional $Q[\chi]$ become linear and
quadratic combinations of the arbitrary coefficients $\{ \tilde \chi^a_m \}$,
\begin {eqnarray}
    \Big( \chi_\ij, \, \S_\ij \Bigr)
    &=&
    \sum_{a,m} \tilde \chi^a_m \, \tilde S_m^a \,,
\\
    \Big( \chi_\ij, \, \C \chi_\ij \Bigr)
    &=&
    \sum_{a,m}
    \sum_{b,n}
    \tilde \chi^a_m \, \tilde C_{mn}^{a\,b} \, \tilde \chi^b_n \,.
\label{eq:def_ctilde}
\end {eqnarray}
The matrix elements $\tilde C_{mn}^{a\,b}$ may be regarded as forming
a square matrix $\tilde C$ of dimension $N_s K$,
while the components $\{ \tilde S^a_m \}$ and $\{ \tilde \chi^a_m \}$
comprise vectors $\tilde S$ and $\tilde \chi$ of the same size.%
\footnote
    {%
    As discussed in Ref.~\cite {AMY1},
    when computing transport coefficients in a plasma with vanishing
    chemical potentials,
    charge conjugation symmetry relates and constrains the
    departures from equilibrium for different species.
    For a QCD-like theory where all matter fields are the same type
    ({\em e.g.}, fundamental representation fermions),
    the net effect is that there is only a single independent
    departure from equilibrium in the C-odd, $\ell = 1$ channel
    relevant for flavor diffusion,
    while there are two independent functions (fermion and gauge boson)
    in the C-even, $\ell = 2$ channel relevant for shear viscosity.
    Hence the actual size of the matrix $\tilde C$ one needs to deal with
    is just the basis size $K$ for diffusion,
    or $2K$ for shear viscosity.
    }
The functional $Q[\chi]$ restricted to this subspace is
$
    \tilde Q[\tilde \chi] =
    \tilde\chi^\trans \tilde S - \half
    \tilde\chi^\trans \tilde C \tilde\chi
$.
The extremum of $\tilde Q[\tilde \chi]$ 
occurs at $\tilde \chi = \tilde C^{-1} \tilde S$,
and
\begin {equation}
    \tilde Q_{\rm max}
    = \half \, \tilde\chi^\trans \tilde S
    = \half \, \tilde S^\trans \tilde C^{-1} \tilde S \,.
\label {eq:Qmax2}
\end {equation}
Note that the variational estimate $\tilde Q_{\rm max}$
gives a lower bound on the true extremum $\Qmax$,
and a nested sequence of variational estimates must converge
monotonically upward to the true result.

The functional $Q[\chi]$ is most sensitive to the behavior of $\chi^a(p)$
for momenta near $T$.
Having factored $f^a_0 [1 \pm f^a_0]$ out of the first order correction
to local equilibrium, as done in Eq.~(\ref {eq:f1}), one may show%
\footnote
    {%
    In the absence of $1 \lra 2$ processes, the asymptotic large and
    small $p$ behaviors are both proportional to $p^\ell$.
    Because of LPM suppressed
    $1 \lra 2$ processes, the large $p$ behavior is modified to 
    $p^{\ell - \half}$.  However, $2 \lra 2$ processes are much more
    efficient so this behavior only sets in at very large momenta
    which are irrelevant for transport coefficients.
    The soft region is also modified by $1 \lra 2$ processes,
    leading to $p^{\ell-1}$ behavior.  This only occurs for $p < \mD$,
    and is an accident of our
    including finite Debye screening masses in $2 \lra 2$ processes but
    treating $1 \lra 2$ processes as perfectly collinear.  
    } 
that the resulting functions $\chi^a(p)$ grow 
no faster than $p^\ell$ as $p \to \infty$.
Also, $\chi^a(p)$ vanishes at least as fast as $p^{\ell - 1}$ at small $p$.
Consequently, one reasonable choice of basis functions is
\begin{equation}
\phi^{(m)}(p) = \frac{ p^{\ell-1} \, (p/T)^{m-1}}{(1+p/T)^{K-2}} \, , 
	\quad m=1,\ldots,K \, .
\label{eq:phi_choice}
\end{equation}
These functions are not orthogonal; they do not need to be.  
As noted below, choosing strictly positive basis functions
improves the accuracy of numerical integrations.
These basis functions span nested subspaces and, as $K \to \infty$,
the basis set becomes complete.%
\footnote
    {%
    For each $K$, our functions $\phi^{(m)}$ are linear combinations of
    the first $K$ functions in the sequence
    $\chi_n(p) = p^{\ell-1} (1{+}p/T)^{1-n}$, $n = 0, 1,\ldots\,$.
    Therefore, as we increase $K$, the span of our functions strictly
    increases.
    One may easily show
    that the functions $\{ \chi_n \}$ form a complete basis in
    $L^2({\mathbb R_+,d\mu})$ with measure
    $d\mu = w(p) \, p^2 \, dp$
    and $w(p)$ any weight function which is
    real, positive, smooth, bounded, and 
    falls faster than $p^{-2l-4}$ as $p \to \infty$.
    The most natural weight function for our application is
    $w(p) = f_0(p)[1{\pm} f_0(p)]$.
    } 
that we expect the
large $K$ limit of the finite $K$ variational extremum to give the
extremal value in the 
full space of allowed $\chi$.

Evaluating matrix elements of the
linearized $2\lra2$ collision operator (\ref {eq:C22})
requires performing an eight dimensional integral (after accounting
for energy and momentum conservation).
By suitably choosing variables, as discussed in Appendix \ref {app:22},
three of these integrals represent overall rotations in momentum space
and are trivial, and one remaining angular integral may be performed
analytically.
This leaves a non-trivial four dimensional integral which must be
evaluated numerically for each pair of basis functions.
It is computationally challenging to perform these
numerical integrations both efficiently and accurately.
Use of a basis of strictly positive, non-orthogonal functions,
such as the set (\ref {eq:phi_choice}),
prevents cancellations between different regions of the integration
which would otherwise degrade the accuracy of the numerically integration.
Of course, one does not want to use a basis which causes the
resulting matrix $\tilde C$ to be so ill-conditioned that its
inversion in formula (\ref {eq:Qmax2}) becomes a problem.
The choice of the functions (\ref {eq:phi_choice}) was motivated
by the need to balance these two conflicting goals.

Due to the collinearity of the $1\lra2$ processes,
computing matrix elements of this part
of the linearized collision operator (\ref {eq:C12})
involves only a two dimensional integral.
As noted earlier, however, evaluating the integrand itself
requires the solution of a two-dimensional linear integral equation.
This evaluation is discussed in Appendix~\ref {app:12};
one can either convert the integral equation to a
differential equation/boundary value problem which
can be solved with a shooting method \cite{Gelis2,AGMZ},
or one may again use a finite basis variational approach \cite{AMY3}.

\subsection {Thermal masses}

In the next two sections
we will present results for SU($N$) or U(1) gauge theories
with $\nf$ Dirac fermions in the fundamental representation.%
\footnote
    {%
    In a theory with chiral (Weyl) fermions, such as the standard
    electroweak theory (ignoring Yukawa interactions),
    each Weyl fermion contributes half as much as a
    Dirac fermion; so for instance, SU(2) theory with 12 left handed
    doublets behaves the same as SU(2) theory with 6 Dirac doublets.
    }
Because the weak coupling behavior of transport coefficients
in hot gauge theories is not given by a simple power series in $g$,
leading order results cannot be presented just by reporting
a coupling-independent value of the leading coefficient.
Rather, leading order results contain non-trivial dependence
on the ratio of the effective thermal mass for hard gauge bosons
to the temperature, $\mg/T$, and the analogous ratio involving
the effective thermal mass for hard fermions, $\mf/T$.
These masses depend on the gauge coupling and on the particle content
of the theory;
at leading order,
\begin{eqnarray}
    \mg^2 &=& {\textstyle \frac 16} \left( \ca + \nf \cf \, \frac{\df}{\da}
                                    \right)
	    \> g^2 \, T^2 
	\equiv {\textstyle \frac 12}\, \mD^2
	\, ,
\label{eq:mD_value}
\\[6pt]
    \mf^2 &=& {\textstyle \frac1{4}} \, \cf \> g^2\, T^2 
	\equiv 2 \mQ^2
	\, .
\label{eq:mQ_value}
\end{eqnarray}
As indicated here, the thermal masses for hard excitations,
$\mg$ and $\mf$, differ by factors of $\sqrt 2$
from the more commonly used Debye mass $\mD$ or thermal quark mass
$\mQ$ (which is the thermal energy of a fermion with zero momentum).
We will generally use the latter variables below.
Here, $\df$ and $\da$ denote the dimensions of the fundamental
and adjoint representations, respectively, while $\cf$ and $\ca$ are
the corresponding quadratic Casimirs.%
\footnote
    {%
    For U(1), $\df=\da=\cf=1$ and $\ca=0$.
    \\For SU(2), $\df = \ca = 2$, $\cf = 3/4$, and $\da = 3$.
    \\For SU(3), $\df = \ca = 3$, $\cf = 4/3$, and $\da = 8$.
    \\For SU($N$), $\df = \ca = N$, $\cf=(N^2{-}1)/(2N)$, and $\da = N^2{-}1$.
    }
Note that the ratio of (leading-order) thermal masses, $\mg/\mf$,
is independent of the
gauge coupling and only depends on the particle content of the theory.

\section {Leading order diffusion and shear viscosity in QCD}
\label{sec:leading-order}

\begin{figure}
\includegraphics[scale=1.00]{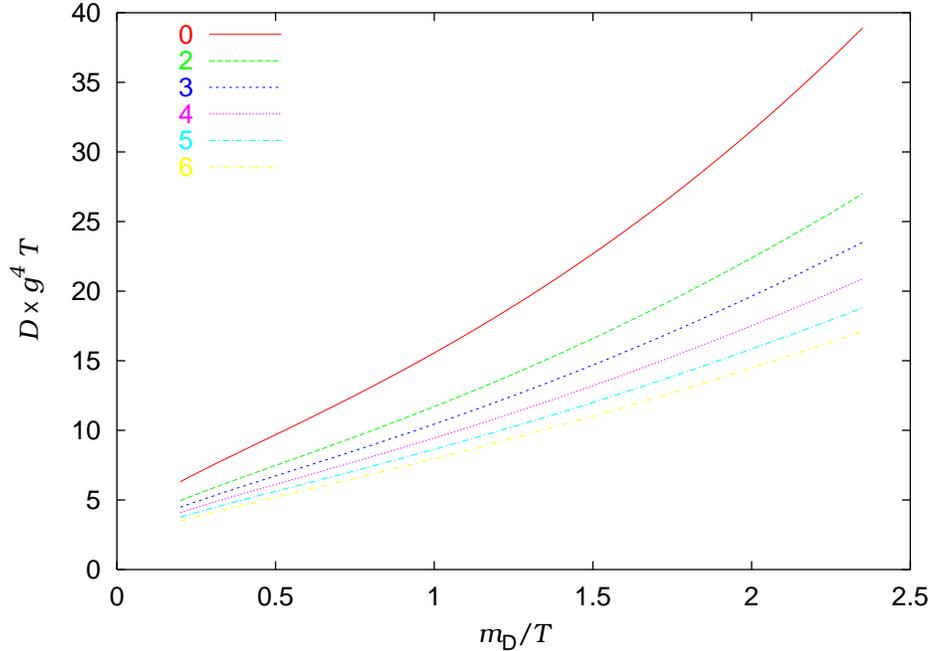}
\caption{%
    Leading-order value of the quark
    flavor diffusion constant $\Dq$, multiplied by $g^4 T$,
    plotted as a function of $\mD/T$.
    The different curves show the result for SU(3) gauge theory
    with 0 to 6 flavors of quarks.
    The $\nf = 0$ curve is the result when one artificially
    neglects scattering of quarks off other quarks,
    and only includes scattering of quarks off gluons
    (in which case the result is independent of $\nf$).
    \la{fig:LO-diffusion}
    }
\end{figure}

\begin{figure}
\includegraphics[scale=1.00]{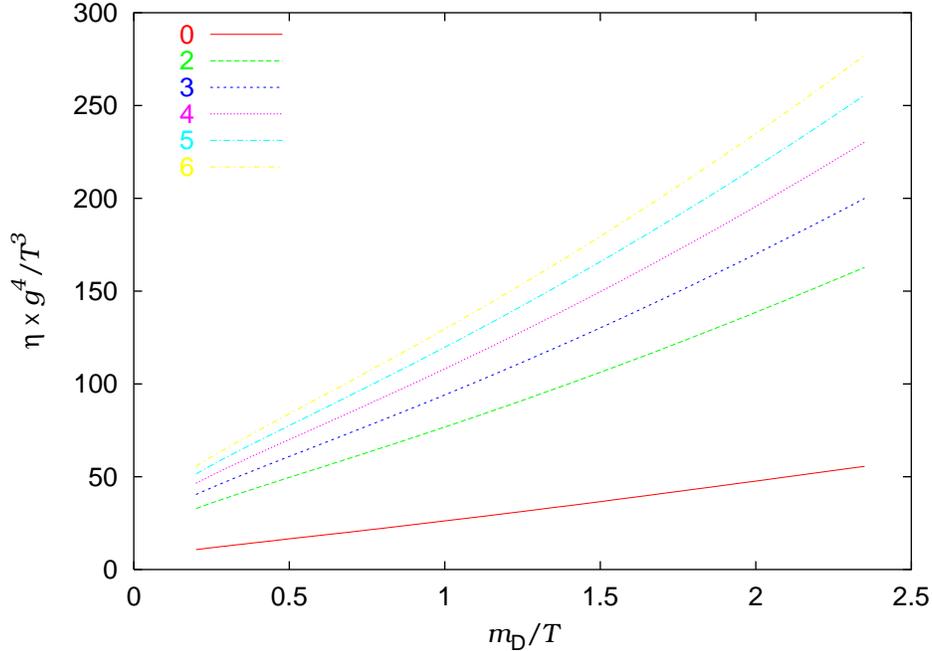}
\caption{%
    Leading-order value of the
    shear viscosity $\eta$, multiplied by $g^4/T^3$, as a function
    of $\mD/T$.
    The different curves show the result for SU(3) gauge theory
    with 0 to 6 flavors of fermions.
    \la{fig:LO-shear}
    }
\end{figure}

In Figure \ref {fig:LO-diffusion} we plot the fermion flavor diffusion
constant $\Dq$ (multiplied by $g^4 T$) as a function of $\mD/T$
for QCD with several different values of $\nf$.
Figure \ref {fig:LO-shear} shows the shear viscosity,
multiplied by $g^4/T^3$,
as a function of $\mD/T$ for the same set of theories.
The results in these figures were computed using four basis functions
of the form (\ref {eq:phi_choice}).
Truncation error due to the use of a finite basis set is
smaller than one part in $10^4$, or
smaller than the width of the lines in Figs.~\ref {fig:LO-diffusion}
and \ref {fig:LO-shear}.  Errors dues to the finite numerical integration
precision are also smaller than the widths of the lines.
If only two basis functions are used, then errors of about 1\% result.


\begin{figure}[tbh]
\includegraphics[scale=0.55]{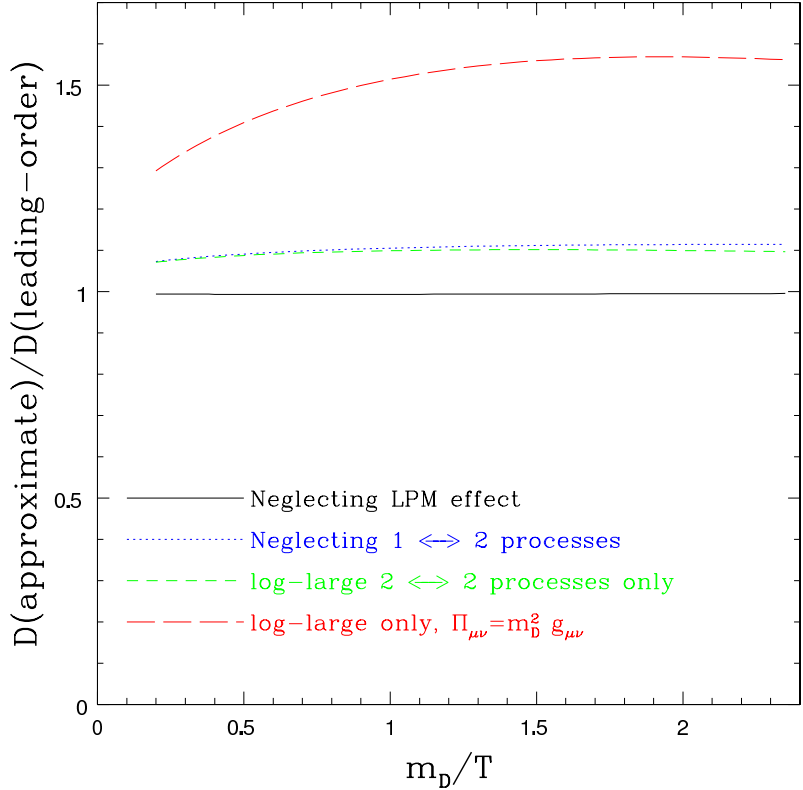}
\caption{
\label{fig:drop_things} 
Sensitivity of the quark diffusion constant in three flavor QCD
to various simplifying approximations.
Each curve shows the ratio of the answer with the indicated
approximations made, to the full leading order answer.}
\end{figure}

\subsection {Quality of various simplifying approximations}

To illustrate the relative importance of different parts of the collision term,
and to examine the sensitivity of results to the correct inclusion
of thermal self-energies,
we plot in Figure~\ref{fig:drop_things} the relative change
in the quark diffusion constant of three flavor QCD
resulting from various (over)simplifications.
The different levels of damage are as follows.
\begin{enumerate}
\item
    Neglect LPM corrections to the rate of $1 \lra 2$ processes.
\item
    Neglect collinear $1 \lra 2$ processes altogether.
\item
    Neglect $1\lra2$ processes and
    those $2\lra2$ terms which do not contribute at leading-log order.
    In other words, include only the underlined terms in the $2\lra2$
    matrix elements shown in Table~\ref{table:mat_elements} in
    Appendix~\ref {app:22}.
\item
    Drop all terms which do not contribute at leading-log order,
    and then replace the correct momentum-dependent HTL self-energies in the
    exchanged propagator by just the Debye mass or thermal quark mass.
    That is, replace $1/t^2$ by $1/(t {-} \mD^2)^2$ in gauge boson
    exchange diagrams, and replace $1/t$ by $t/(t{-}\mQ^2)^2$ in
    fermion exchange diagrams.
\end{enumerate}
None of these approximations are correct at leading order in $g$, but
some are far more numerically significant than others.
As one may see from Fig.~\ref {fig:drop_things},
neglecting the LPM effect makes quite a small change in the results, as
does dropping $2 \lra 2$ terms which do not contribute at leading-log order.%
\footnote
    {%
    Both of these approximations actually overestimate the collision term.
    The LPM effect suppresses scatterings, so its neglect increases 
    the collinear splitting rate.
    And, in SU(3) gauge theory,
    dropping the subdominant $2 \lra 2$ terms increases the collision
    rate because the most important neglected term is an 
    interference term which is negative in a non-Abelian theory.
    }
However, the collinear $1 \lra 2$ processes are
important at the 10\% level,
while the approximation sometimes used in the literature of replacing
the bosonic self-energy by just the Debye mass is quite poor
and results in errors at the 50\% level.

\subsection {Sensitivity to higher-order corrections}

The leading-order results have relative corrections of order $g$,
and hence are only reliable when $\mD/T$ is sufficiently small.
Of course, for a given level of precision,
specifying just how small is ``small enough''
is not possible without knowing more about the
actual size of sub-leading corrections.

It should be emphasized that our leading order results
do depend on specific choices which were made in defining the
linearized collision operator $\C$.
We would obtain somewhat different numerical results
if we had included thermal self-energies on internal lines of diagrams
even when this was not strictly necessary,
if we had used full one-loop self-energies instead of their HTL
approximation (valid for momenta small compared to $T$),
if we had included thermal corrections to the on-shell dispersion
relations in $\twotwo$ processes,
or if we had not approximated the nearly collinear $\onetwo$ processes
as strictly collinear.
All of these effects produce relative changes suppressed by at least
one power of $g$, so we are justified in neglecting them.
However, handling any of these issues differently can lead to
different, but equally valid, formally leading-order results.

    Examining the sensitivity of results to changes in the
precise definition of the linearized collision operator $\C$
(which are equally valid at leading-order) is one way to get
a ``hint'' as to the likely size of some actual $O(g)$ sub-leading effects.
Fig.~\ref {fig:differences} shows results for the diffusion constant
in three flavor SU(3) gauge theory,
computed using several different, but equally valid at leading-order,
definitions of the linearized collision operator.

We have only considered
modifications of the way in which the thermal gauge field self-energy is
introduced in $t$ or $u$ channel gauge boson exchange diagrams.
We focus on these modifications
because these diagrams numerically dominate the collision term, and
because such modifications are easy to study. 
Our ``standard'' choice
is discussed in Appendix A; it consists of writing $t$ (or $u$)
channel exchange diagrams as the analogous result for scalar quarks,
computed with the HTL self-energy included in the gauge boson propagator,
plus a spin-dependent infrared safe remainder in which the HTL self-energy
may be neglected.
Our first alternative is to instead write $t$ or $u$ channel gauge boson
exchanges as the result for fermion-from-fermion scatterings,
plus a (different) IR safe piece.
Expressions for fermion-from-fermion scattering (with the HTL gauge boson
self-energy included) may be found in Appendix B of Ref.~\cite{largeN}.

\begin{figure}
\includegraphics[scale=0.55]{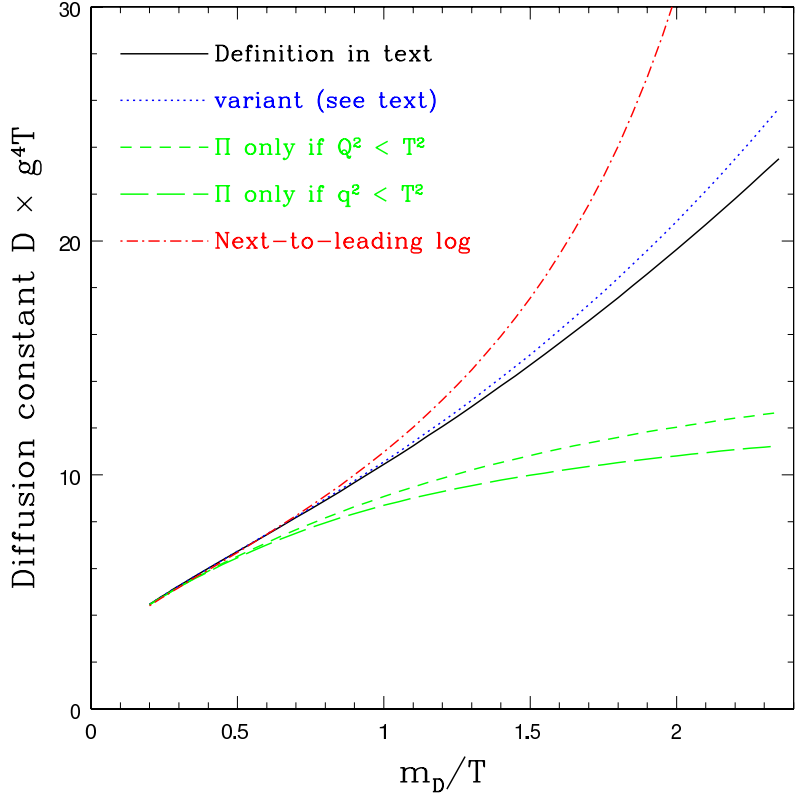}
\caption
    {%
    Leading-order results for the
    flavor diffusion constant $\Dq$, multiplied by $g^4 T$,
    as a function of $\mD/T$
    in three flavor SU(3) gauge theory.
    Each curve is computed using a different, but equally valid, 
    leading-order definition of the effective kinetic theory.
    The solid line shows the result of the implementation discussed
    in detail in Appendix \ref {app:22}, in which $t$ or $u$ channel
    matrix elements are written as an HTL-corrected scalar quark
    contribution plus an IR-safe spin-dependent remainder.
    The dotted line is the result of the analogous procedure using fermionic,
    rather than scalar, scattering as the template for $t$ channel gauge boson
    exchange.  The dashed lines show what happens if the hard thermal
    loop self-energy is only included in gauge boson exchange lines when
    the exchange momentum $Q$ satisfies $Q^2 < T^2$,
    or alternatively $q^2 < T^2$.
    \la{fig:differences}
    }
\end{figure}

The inclusion of hard thermal loop self-energies on $t$ or $u$ channel
exchange lines is only justified for soft exchange momentum,
since otherwise the HTL approximation to the full one-loop self-energy
is invalid.
For large exchange momentum,
inserting the HTL self-energy is no more correct than neglecting
the self-energy altogether,
but in this regime self-energy corrections are $O(g^2)$ effects.
So in a leading-order treatment,
one valid approximation is to multiply
the HTL self-energy by the step function $\Theta(T^2 {-} Q^2)$,
with $Q$ the exchanged 4-momentum.%
\footnote
    {%
    We use a $({-}{+}{+}{+})$ spacetime metric,
    so $Q^2 > 0$ for a spacelike momentum exchange.
    }
Alternatively, one may multiply the HTL self-energy by $\Theta(T^2 {-} q^2)$,
with $q = |\q|$ the spatial 3-momentum transfer.

These various possibilities are compared in Fig.~\ref{fig:differences}.
The differences between these curves are all formally at most $O(g)$.
The width of the band of results, for a particular value of $\mD/T$,
may be viewed as a guess as to the size of certain types of
actual sub-leading corrections.
The figure also includes the next-to-leading logarithmic approximation
from the next section, for comparison.
One sees that the different, but equally valid,
leading-order implementations agree to within 15\% provided
$\mD/T \le 0.8$.
This suggests that the leading-order results might have
about as large a range of utility as one could have reasonably hoped for;
certainly one should not expect a leading-order weak-coupling analysis
to be accurate when $\mD \ge T$.

An obvious question is whether there is some natural way to
define a unique leading-order result which would systematically drop
all $O(g)$ corrections while retaining the full dependence on $1/\ln g^{-1}$.
This is directly related to the summability of the asymptotic expansion
in $1/\ln g^{-1}$, and will be discussed in Appendix~\ref{app:converge}.

\section{Expansion in inverse logs}
\label{sec:log-expansion}

As discussed in Refs.~\cite{BMPRa,Heiselberg,AMY1},
part of the linearized collision operator $\C$
contains a logarithmic enhancement
proportional to $\ln(T/\mD) \sim \ln (1/g)$ arising
from $t$ (or $u$) channel $2\lra2$ exchange processes
with momentum transfer between the Debye screening scale
$\mD$ and $T$.
One may separate the full leading-order collision operator $\cal C$ into
a logarithmically enhanced piece, which we will denote by $\CLL$,
and a remainder $\delta\C$,
\begin{equation}
    \C = \CLL + \delta \, \C \, .
\end{equation}
Expanding the inverse collision operator appearing in
$\Qmax$ [{\em c.f.}, Eq.~(\ref {eq:Qmax1})]
in powers of $\delta \C$ will generate 
an asymptotic expansion of transport coefficients in powers of
$1/\ln g^{-1}$.

To make this separation precise, let $\overline\C \equiv \C/(g^4 T)$
and note that this rescaled operator
is dimensionless and depends on $g$
only through the ratios $\mD/T$ and $\mQ/T$.
Equivalently, since the (leading-order) ratio $\mQ/\mD$ has a fixed
value in a given theory, the $g$ dependence of $\overline\C$ may be regarded
as arising only through dependence on $\mD/T$.
This dependence may be isolated by introducing a separation scale
$q_*$ satisfying $\mD \ll q_* \ll T$
and splitting the relevant $t$ and $u$ channel exchange parts of
matrix elements of $\overline\C$ into contributions from exchange momentum
less or greater than $q_*$,
\begin {equation}
    \Bigl( \chi_\ij, \, \overline\C \, \chi_\ij \Bigr)
    =
    \Bigl( \chi_\ij, \, \overline\C_{[q < q_*]} \, \chi_\ij \Bigr)
    +
    \Bigl( \chi_\ij, \, \overline\C_{[q > q_*]} \, \chi_\ij \Bigr) \,.
\end {equation}
When $q > q_*$, one may safely set $\mD = \mQ = 0$.
This contribution is therefore independent of $\mD$ and $\mQ$,
up to corrections subleading in $g$.
When $q < q_*$, one may expand in $q$ and keep only the first nontrivial order.
(This is safe because in all but a $g^2$ suppressed part of the
integration domain,
all other momenta are large compared to $q_*$.)
As far as the $q$ and $\mD$ dependence is concerned,
the result has the form \cite {Heiselberg,AMY1}
\begin {equation}
    \int_0^{q_*} \frac {dq}q \> f\Big(\frac \mD q \Big) \,.
\end {equation}
Rescaling $q \to q/\mD$, the integrand becomes $g$ independent
and all remaining $g$ dependence is isolated in the upper limit of $q_*/\mD$.
Simplifying the integrand at $q = q_*$, using the assumed scale
separation $\mD \ll q_* \ll T$, leads to an explicit form for
\begin {equation}
    \A \equiv \lim_{\mD/T \rightarrow 0} 
	-\mD \frac {\partial \, \overline \C(\mD/T) }{\partial \mD} \,,
\label{eq:partialg}
\end {equation}
given in Ref.~\cite {AMY1}.
For a gauge theory with $\nf$ Dirac
fermions all in the same representation,
\begin{align}
    \left( \chi_\ij, \A \chi_\ij \right)
    = {} &
	\frac{\nf \df \, \cf^2 \, \beta^2}{32 \pi^3}
	\int_0^\infty \!  dp \> p \>
	f_0^f(p) [1{+}f_0^g(p)]
	\left( \vphantom {\chi^{\bar f}}
	    [ \chi^f(p){-}\chi^g(p) ]^2 + [ \chi^{\bar f}(p){-}\chi^g(p) ]^2
	\right)
\nonumber \\ + {} &
	\frac{\da \, \ca \, \beta^3 \, \mD^2}{(2 \pi)^3 \, g^2} \,
	\int_0^\infty \! dp \>
	f_0^g(p)[1{+}f_0^g(p)]
	\left( \vphantom {\chi^{\bar f}}
	    p^2[\chi^g(p)']^2 + \ell(\ell{+}1) [\chi^g(p)]^2
	\right)
\nonumber \\ + {} &
	\frac{\nf \, \df \, \cf \, \beta^3 \,\mD^2}{(2 \pi)^3 \, g^2} \,
	\int_0^\infty \!\! dp \>
	f_0^f(p)[1{-}f_0^f(p)]
	\left( \vphantom {\chi^{\bar f}}
	    p^2[\chi^f(p)']^2 + \ell(\ell{+}1) [\chi^f(p)]^2
	\right. \; \quad &
\nonumber \\ & \hspace*{2.6in} {} + 
	\left. \vphantom {\chi^{\bar f}}
	    p^2[\chi^{\bar f}(p)']^2 + \ell(\ell{+}1) [\chi^{\bar f}(p)]^2
	\right) . \!
\label{eq:C_LL}
\end{align}
Here 
$\chi^g$, $\chi^f$, and $\chi^{\bar f}$ are the
departures from equilibrium
for gauge bosons, fermions, and anti-fermions, respectively,
and primes denote derivatives.
The above form assumes that departures from equilibrium are
fermion flavor independent,
as is appropriate for computing shear viscosity or baryon number diffusion.
The prefactor of $\beta^2 \mD^2/g^2$ in two of the terms is just a compact way
of writing $\frac 13 (\ca + \nf \cf \, \df/\da)$.
The first integral comes from
fermion exchange diagrams whose infrared behavior
is regulated by $\mQ$, while the latter two integrals arise from
gauge boson exchange diagrams whose IR
behavior is regulated by $\mD$.

Since the operator $\A$ is itself $\mD$ independent,
the definition (\ref{eq:partialg}) implies that the limit
\begin {equation}
    \delta\overline\C (\mu/T)
    \equiv
    \lim_{\mD\to0} \Bigl[\, \overline\C(\mD/T) - \A \, \ln (\mu/\mD) \Bigr]
\label{eq:lim1}
\end {equation}
exists.
Therefore the original linearized collision operator,
up to $O(g)$ corrections,
may be written as
\begin{eqnarray}
    \C &=& \CLL(\mu) + \delta \C(\mu)\, ,
\label{eq:C_pieces}
\\
\noalign {\hbox{where}}
    \CLL(\mu) &\equiv& g^4 T \> \A \, \ln ({\mu}/{\mD}) \,,
\label {eq:CLL}
\\
\noalign {\hbox{and}}
    \delta \C(\mu) &\equiv& g^4 T \> \delta \overline\C (\mu/T)  \,.
\label {eq:delC}
\end{eqnarray}
Here $\mu$ is an arbitrary scale which should be chosen to be $O(T)$
so that $\delta\C(\mu)$ does not contain large logs.

Expanding the inverse collision operator in a geometric series,
\begin {equation}
    \C^{-1}
    =
    \left[ \CLL(\mu) + \delta\C(\mu) \right]^{-1}
    \sim
    \sum_{n=0}^\infty \; (-1)^n \> \CLL(\mu)^{-1}
    \left[ \delta \C(\mu) \, \CLL(\mu)^{-1} \right]^n \,,
\label {eq:C_inv}
\end {equation}
directly gives its asymptotic expansion
in powers of $[\ln (\mu/\mD)]^{-1} \sim 1/\ln g^{-1}$.
Inserting this expansion into
$\Qmax = \half \left( \S_\ij , \, \C^{-1} \S_\ij \right)$
and using the definitions (\ref {eq:CLL}) and (\ref {eq:delC})
yields the inverse log expansion of $\Qmax$
(and hence of transport coefficients),
\begin {equation}
    \Qmax
    \sim
    \frac 1{g^4\,T}
    \sum_{n=1}^\infty \; Q_n(\mu)
    \left( \ln \frac {\mu}{\mD} \right)^{-n} \,,
\label {eq:Qseries}
\end {equation}
with
\begin {equation}
    Q_n(\mu)
    \equiv
    \half \, (-1)^{n-1} \;
    \Bigl( \S_\ij ,\,
    \A^{-1} \left[ \delta \overline\C(\mu) \, \A^{-1} \right]^{n-1}
    \S_\ij \Bigr) \,.
\label {eq:Qn}
\end {equation}
Except for $Q_1$,
the coefficients $Q_n(\mu)$ are $\mu$-dependent.
However, the series is formally $\mu$ independent
in exactly the same way that perturbative series in QCD are
formally renormalization point independent even though
individual $n$-loop contributions do depend on the renormalization point.

To avoid the presence of large logarithms in the coefficients $Q_n(\mu)$,
one must choose $\mu$ to be $O(T)$ parametrically,
but the exact coefficient is not uniquely prescribed.
One somewhat natural choice is to select the value $\mu_*$
for which $Q_2(\mu_*)$ vanishes.
This can be termed the fastest apparent convergence (FAC)
choice of $\mu$ at next-to-leading log order.
Given results for $Q_1$ and $Q_2(\mu)$ at some other value of $\mu$,
\begin{equation}
    \mu_* = \mu \, \exp[-Q_2(\mu) / Q_1] \, .
\end{equation}

To evaluate the coefficients $Q_n(\mu)$ of the expansion (\ref {eq:Qseries}),
we use exactly the same finite basis set approach described in section
\ref {sec:vary}.
The linear operators $\A$ and $\delta \overline\C$ are replaced
by their matrix representations $\tilde A$ and $\widetilde {\delta C}$
in the finite basis set,
and 
\begin {equation}
    \tilde Q_{\rm max}
    \sim
    \frac 1{g^4 \, T}
    \sum_{n=1}^\infty \; \tilde Q_n(\mu)
    \left( \ln \frac {\mu}{\mD} \right)^{-n} \,,
\label {eq:Qseries2}
\end {equation}
with
\begin {equation}
    \tilde Q_n(\mu)
    =
    \half (-1)^{n-1} \;
    \tilde S^\trans
    \tilde A^{-1} \left[ \widetilde {\delta \C}(\mu) \,
    \tilde A^{-1} \right]^{n-1}
    \tilde S \,.
\label {eq:Qn2}
\end {equation}

The limit (\ref{eq:lim1}) defining $\delta \overline \C(\mu)$
is performed numerically by evaluating each matrix element of $\overline \C$
at several small values of $\mD$,
subtracting off the leading log piece
(whose matrix elements are easy to evaluate),
and then extrapolating to vanishing Debye mass.
We find that this extrapolation is quite well behaved,
although the numerical integrals at small $\mD$ or $\mQ$ become
rather demanding.  

Using the basis functions (\ref {eq:phi_choice}),
we previously found \cite{AMY1} that
the fractional difference between $Q_1$ and its finite basis approximation
$\tilde Q_1$ is less than $10^{-5}$ with 4 basis elements and less
than $10^{-6}$ using six.
The higher $Q_n$ are more sensitive both to basis size and to numerical
integration errors: $Q_2$ can be reliably determined with 4 basis
functions, and $Q_3$
and $Q_4$ can be found with reasonably small errors using 6 to 8 basis
functions, but higher moments
become rapidly more difficult to evaluate, showing poor convergence with
basis set size and high numerical integration error sensitivity when
the basis sets become very large.
Consequently we have been unable to go very deep in the $Q_n$ series.

\begin {table}

\tabcolsep 10pt
\begin {tabular}{c|c|cc|cc}
      \smash{\lower 10pt\hbox{group}}
    & \smash{\lower 10pt\hbox{$\nf$}}
    & \multicolumn{2}{c|}{flavor diffusion}
    & \multicolumn{2}{c}{shear viscosity}
\\
    & & $D_1$ & $\mu_*/T$ & $\eta_1$ & $\mu_*/T$
\\\hline
    & 1 & 47.089 & 2.469 & 188.38 & 5.007
\\
    \smash{\raisebox{10pt}{U(1)}}  & 2 & 30.985 & 3.013 & 120.28 & 4.418
\\\hline
    SU(2) & 6 & 21.283 & 3.123 & 200.533 & $2.927$
\\\hline
    & 0 & 16.060 & 2.699 & 27.126 & 2.765
\\
    & 2 & 12.999 & 2.887 & 86.47  & 2.954
\\
    & 3 & 11.869 & 2.949 & 106.66 & 2.957
\\
    SU(3) & 4 & 10.920 & 2.997 & 122.96 & 2.954
\\
    & 5 & 10.111 & 3.035 & 136.38 & 2.947
\\
    & 6 & 9.414  & 3.065 & 147.63 & 2.940
\\
    & $\infty$ & $136.76/\nf$ & 3.155 & 274.83 & 2.733
\end {tabular}
\caption
    {
    Values of the leading-log coefficient $D_1$ and $\eta_1$ together
    with the value of $\mu_*/T$,
    for the case of fermion flavor diffusion and shear viscosity
    in theories with the indicated gauge group and $\nf$ Dirac fermions
    in the fundamental representation.
    The large $\nf$ result in the last line 
    is from Ref.~\cite{largeN}, and shows that
    $\nf=6$ is still a long ways from the large $\nf$ limit.
    The values shown for $\nf{=}0$ flavor diffusion represent the
    results one would obtain if diffusing quarks could only scatter
    off gluons, and not off other quarks.
    \label {table1}
    }
\end {table}

Table \ref {table1} shows results for
$\mu_*$ and the first coefficient of the inverse log expansion,
for the case of fermion flavor diffusion and shear viscosity in theories with
various gauge groups and the indicated number $\nf$ of Dirac fermion
flavors in the fundamental representation.
Specifically, we show the first coefficients $D_1$ and $\eta_1$ of the
series%
\footnote{
  The diffusion constant $D$ and shear viscosity $\eta$ are
  related to their respective $Q_{\rm max}$'s as shown in Eq.~(\ref{eq:anyQ}).
  Explicitly,
  $\eta = \frac2{15} \, Q^{(\eta)}_{\rm max}$
  and $D = \frac 23 Q^{(D)}_{\rm max}/(\nf T^2)$ with
  $Q^{(D)}$, in this normalization,
  being associated with {\it total}\/ quark number.
  Hence, the inverse log expansion coefficients in
  Eqs.~(\ref {eq:Dexp}) and (\ref {eq:etaexp}) are related to the
  previous coefficients $Q_n(\mu)$ via
  $D_n(\mu) = \frac 23 Q^{(D)}_n(\mu)/(\nf T^2)$ and
  $\eta_n(\mu) = \frac 2{15} Q^{(\eta)}_n (\mu)/ T^4$.
}
\begin {equation}
    D
    \sim
    \frac 1{g^4 \, T}
    \sum_{n=1}^\infty \; D_n(\mu)
    \left( \ln \frac {\mu}{\mD} \right)^{-n}
\label {eq:Dexp}
\end {equation}
and
\begin {equation}
    \eta
    \sim
    \frac {T^3}{g^4}
    \sum_{n=1}^\infty \; \eta_n(\mu)
    \left( \ln \frac {\mu}{\mD} \right)^{-n} .
\label {eq:etaexp}
\end {equation}
Together with the corresponding values of $\mu_*/T$, these numbers determine
the next-to-leading-log (NLL) approximation to the respective
transport coefficients,
\begin {eqnarray}
    D_{\rm NLL} &=&
    \frac 1{g^4 \, T} \left[ \frac {D_1}{\ln (\mu_*/\mD)} \right]
\\\noalign{\hbox{and}}
    \eta\strut_{\rm NLL} &=&
    \frac {T^3}{g^4} \left[ \frac {\eta_1}{\ln (\mu_*/\mD)} \right] .
\end {eqnarray}
Note that in SU(3) gauge theory,
$\mu_*/T$ is quite close to 3
for both transport coefficients,
regardless of the number of fermion flavors.

We have computed further terms in the inverse log expansion
in the case of SU(3) gauge theory with 3 fermion flavors.
For flavor diffusion, we find
\begin {equation}
    D_3(\mu_*) = 2.436(2) \,,\qquad
    D_4(\mu_*) = -0.11(1) \,,\qquad
    D_5(\mu_*) = 1.7(1)   \,,\phantom {00}
\end {equation}
while for shear viscosity
\begin {equation}
    \eta_3(\mu_*) = 27(1) \,,\phantom{.000}\qquad
    \eta_4(\mu_*) = 6(5)  \,,\phantom{-0.00}\qquad
    \eta_5(\mu_*) = 100(100) \,.
\end {equation}
[And $D_2(\mu_*) = 0 = \eta_2(\mu_*)$, by our definition of $\mu_*$.]
The third-order coefficients $D_3(\mu_*)$ and $\eta_3(\mu_*)$ are
roughly one quarter the size of $D_1$ and $\eta_1$, respectively.
The next order coefficients $D_4(\mu_*)$ and $\eta_4(\mu_*)$ are yet smaller,
but subsequent coefficients appear to grow.

\begin{figure}
\!\!\!\includegraphics[scale=0.42]{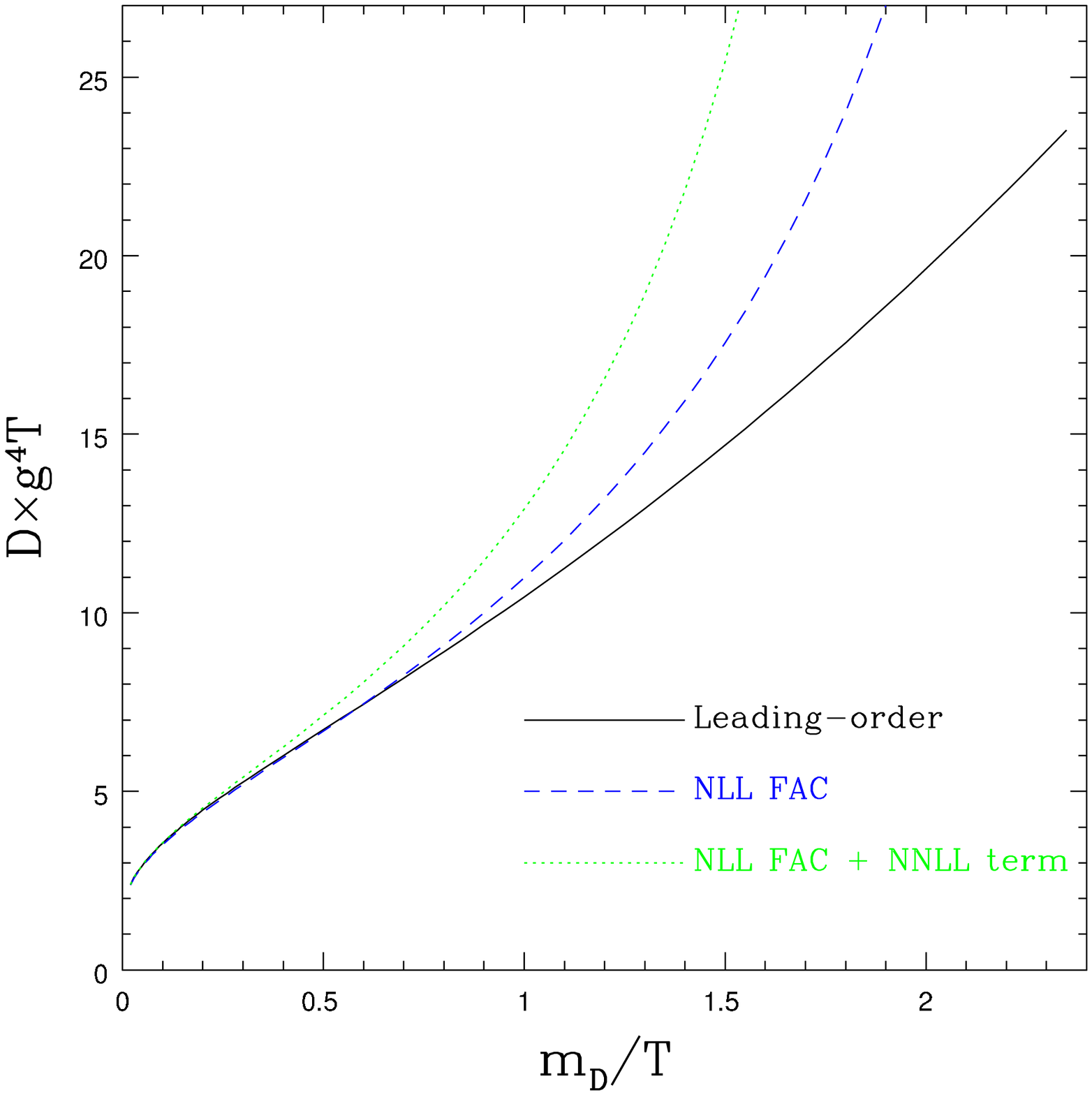}\quad
\includegraphics[scale=0.42]{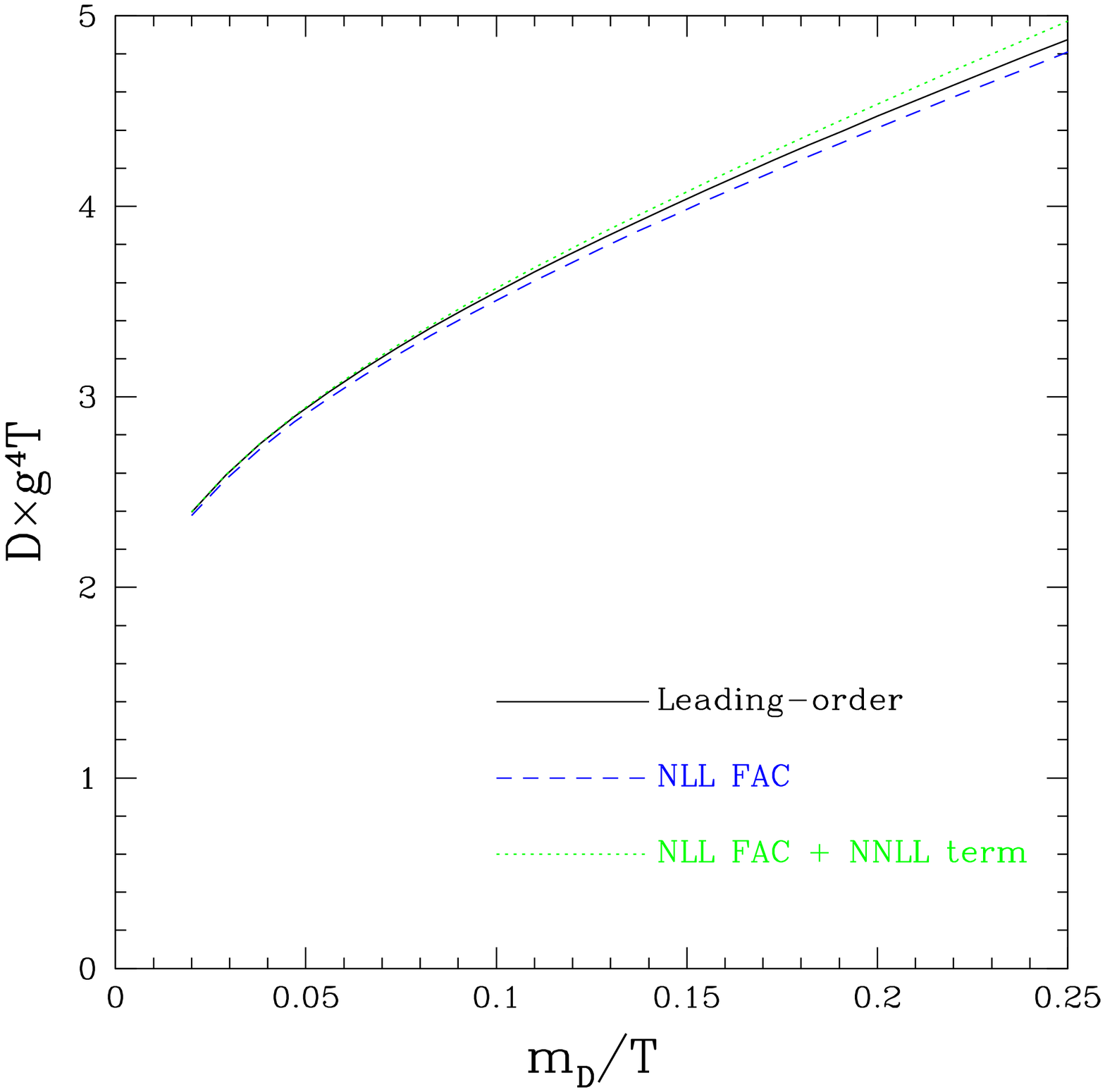}\!\!\!
\caption{%
    Leading-order results for the flavor diffusion constant
    (multiplied by $g^4 T$) in three flavor QCD
    compared to the expansion in inverse powers of $\ln (\mu_*/\mD)$
    truncated at second and third order.  Right panel: zoom-in on the
    small $m_D/T$ region, showing that the third order truncation
    of the inverse-log expansion can be an improvement over the
    second order result, but only for $\mD/T \le 0.2$.
    \la{fig:nll}
    }
\end{figure}

The behavior of the expansion in inverse powers of $\ln(\mu_*/\mD)$,
truncated at second or third order, is compared
to the full leading-order result in Figure~\ref {fig:nll},
for the case of flavor diffusion in three flavor QCD.
As the figure makes clear, the next-to-leading-log result
is remarkably close to the full leading order result out to
$\mD/T = 1$, but going beyond second order in the
inverse log expansion has very little practical utility.

Appendix \ref{app:converge} discusses the asymptotics of the
inverse log expansion and proves that this expansion is, in fact,
a typical asymptotic expansion with zero radius of convergence.
An argument is also given suggesting that this
asymptotic expansion is not Borel summable.
As is well known,
the presence of singularities in the Borel transform
on the positive real axis generates
ambiguities in the inverse Borel transform.
In the typical case of a power series in $g^2$,
this irreducible ambiguity is non-perturbative,
behaving as $\exp (-c/g^2)$ where $c$ is the
location of the singularity of the Borel transform
nearest to the origin on the positive real axis.
(See, for example, Refs.~\cite {Borel1,Borel2}.)
In the present case where the expansion parameter is an inverse log
of the coupling,
a singularity in the Borel transform instead indicates an
inherent ambiguity in the Borel sum of the asymptotic series
which is a power of coupling.
The estimate of appendix \ref{app:converge} suggests that
this is an $O(g^5)$ ambiguity.

\section {Electrical conductivity}
\label {sec:elec-cond}

We have also computed the electrical conductivity for a high temperature plasma
of leptons, or leptons plus quarks.
As mentioned earlier, the electrical conductivity is related to the
diffusion constants of charged species via the Einstein relation
(\ref{eq:einstein}).
In a plasma of leptons plus quarks,
we make the same $\alphaEM^2 \ll \alphas^2$
approximation used in previous work
\cite{BaymHeiselberg,AMY1}:
we neglect the electric current directly carried by quarks,
and only compute the charged lepton diffusion constant.
Because quarks undergo efficient QCD scattering
(as compared to QED scatterings), their departure from equilibrium
in the presence of an electric field is negligible compared to
that of charged leptons, and hence so is their contribution to the
electric current.
This approximation amounts to the neglect of relative
$O(\alphaEM^2 / \alphas^2)$ corrections to the conductivity.
Quarks remain relevant, however, as excitations off of which charged
leptons can scatter.
Of course, for plasmas containing quarks we still require $\alphas \ll 1$,
so that quarks may be treated as nearly free massless excitations.

\begin{table}

\tabcolsep 10pt
\begin{tabular}{c|c|c|c|c|c|c} 
\smash{\lower 10pt\hbox{leptons}}
& \smash{\lower 8pt\hbox{quarks}}
& \smash{\lower 8pt\hbox{$\sum q^2$}}
& \smash{\lower 8pt\hbox{$\sigma_1$}}
& \smash{\lower 8pt\hbox{$\mu_*/T$}}
& \multicolumn{2}{c}{$\sigma \times e^2/T$}
\\ &&&&& \multicolumn{1}{c}{NLL FAC}&leading order%
\\ \hline
$e$          & $-$         & 1    & 15.696 & 2.470 & 5.928 & 5.9700 \\
$e,\mu$      & $-$         & 2    & 20.657 & 3.013 & 8.262 & 8.2996 \\
$e,\mu$      & $u,d,s$     & 4    & 12.287 & 3.268 & 5.498 & 5.4962 \\
$e,\mu,\tau$ & $u,d,s,c$   & 19/3 & 12.520 & 3.306 & 6.208 & 6.1756 \\
$e,\mu,\tau$ & $u,d,s,c,b$ & 20/3 & 11.972 & 3.306 & 6.013 & 5.9769 \\
\end{tabular}
\caption{\label{table:conduct}
Electrical conductivity
in plasmas containing the indicated types of leptons and quarks.
Each entry is relevant for temperatures such that
the listed species are much lighter than $T$
while all other leptons or quarks are much heavier than $T$.
Relative corrections of order $\alphaEM^2/\alphas^2$ are neglected;
see text.
The last two columns compare the next-to-leading-log (NLL FAC) approximation
with the full leading-order result,
both evaluated at the physical value of $\mD$ (using $\alphaEM = 1/137.04$).
Clearly, the NLL FAC approximation works very well for QED.}
\end{table}

For simplicity we have only
analyzed physically relevant combinations of leptons and quarks.
We have evaluated the electrical conductivity $\sigma$
at next-to-leading-log order (NLL FAC)
as well as at full leading order.
In the next-to-leading-log form,
\begin {equation}
    \sigma_{\rm NLL}
    =
    \frac T{e^2}
    \left[ \frac {\sigma_1}{\ln (\mu_*/\mD)} \right] ,
\end {equation}
the Debye mass now refers to the inverse QED screening length
given by
\begin {equation}
    \mD^2 = {\textstyle \frac 13} \, e^2 T^2 \>
    \Bigl( {\textstyle \sum_i} \> q_i^2 \Bigr) \,.
\end {equation}
$q_i$ is the charge assignment of a given species, and the
sum runs over all Dirac fermions.
Instead of presenting plots showing the conductivity as a function of
$\alphaEM$, we have simply set $\alphaEM = 1/137.04$.%
\footnote{
   We use this many significant digits, ignoring the running of the
   coupling,
   for no reason other than to show the precision of the numerics
   and to compare different results.
}
Our results are presented in Table \ref{table:conduct}.
For this quite small value of coupling, one sees that the NLL FAC
treatment agrees with the full leading order result to better
than 1\%.

\section{Conclusion}

\iffrivolous
\begin {verse}

   Well it's time for so long,   \\
   But we'll sing one more song. \\
   Thanks for doing your part, \\
   You sure are smart. \\
   With me, and you, and our friend Blue, \\
   We can do, anything, that we want to do.

\end {verse}
\else

    We have performed complete leading-order calculations of shear viscosity,
electrical conductivity, and fermion diffusivity in QCD and QED.
``Leading-order'' means that all neglected effects are suppressed by
one or more powers of the gauge coupling $g(T)$.
To our knowledge, this is the first time any transport coefficient has
been evaluated with leading-order accuracy in a high temperature gauge theory.
Due to the presence of Coulomb logarithms arising from small angle scattering,
the coefficient of the leading power of $g(T)$ is not a simple number,
as in scalar theories, but rather is a non-trivial function of $\ln(g^{-1})$.

Leading-order results for transport coefficients
may themselves be expanded in powers of $1/\ln(g^{-1})$.
We have explicitly computed both leading and first sub-leading terms
for shear viscosity and quark diffusivity in
U(1), SU(2), and SU(3) gauge theories with various numbers of
fermion fields (as well as several more terms for three flavor QCD).
For QCD, the next-to-leading log result (with the sub-leading term
absorbed by suitably shifting the scale inside the leading log)
was found to be remarkably close to the full leading-order result
as long as $\mD/T \le 1$.
This is a much larger domain of utility than one might have expected.
For these transport coefficients,
we also find that only roughly 10\% errors are made if one neglects
near-collinear $1 \lra 2$ particle splitting processes,
which are considerably more difficult to analyze than $2 \lra 2$ particle
scattering processes.
(However, it should be noted that some transport coefficients which we have not
analyzed, such as bulk viscosity, depend primarily on particle
number-changing processes and so may be expected to depend essentially
on $1\lra2$ processes.)

Because the expansion in inverse powers of $\ln(g^{-1})$ is only
asymptotic, not convergent, as demonstrated in Appendix \ref {app:converge},
we are not able to give a unique, unambiguous prescription for separating
leading-order contributions from higher-order effects.
As discussed in Appendix \ref {app:converge}, it appears that
the inverse log expansion is not Borel summable,
which would imply that no clean separation is possible.
In practice, this means that any specific calculation yielding results
of leading-order accuracy will necessarily include some contributions
from higher-order effects.
However, our examination of several different 
prescriptions for computing leading-order results suggests
that this is not a significant issue for $\mD \lsim 0.8 \, T$.

\begin{figure}
\includegraphics[scale=0.40]{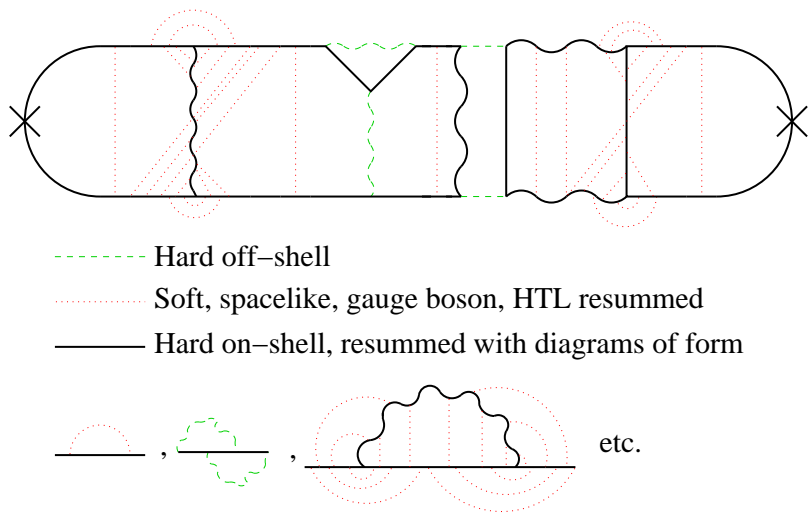}
\caption{\label{fig:awful} Typical diagram needed in the leading-order
evaluation of the shear viscosity in QCD.  The crosses at the left and
right denote $T_{ij}$ (stress tensor) insertions.}
\end{figure}

Our tool for studying transport coefficients has been kinetic theory,
specifically the effective kinetic theory presented in our previous paper
\cite{AMY5}.  As originally shown by Jeon \cite{Jeon},
in the context of weakly-coupled relativistic scalar theories,
it is also possible to compute transport coefficients diagrammatically
starting from the appropriate Kubo formulae involving current-current or
stress-stress correlators.
Such a diagrammatic approach amounts to a complicated way to
derive the appropriate linearized Boltzmann equation specialized
to the particular symmetry channel of interest.
For gauge theories, this diagrammatic approach has been applied,
only at leading logarithmic order, to the
electrical conductivity by Valle Basagoiti \cite{Basagoiti}.
Trying to use a diagrammatic approach
for a complete leading order calculation
would be an enormously more difficult task.
For instance, a typical diagram which we believe
contributes at leading order to the
$\langle T_{ij} T_{kl} \rangle$ correlator, needed for the shear
viscosity, is depicted in Figure~\ref{fig:awful}.
Note that the hard (nearly) on-shell propagators require
self-energy resummations, illustrated at the bottom of the figure,
which go far beyond the HTL approximation.
The one and two loop self-energy contributions shown account for
scattering via $2\leftrightarrow 2$ processes,
while the very complicated self-energy diagram
represents one contribution to the effective $1 \leftrightarrow 2$
splitting process; the reason it needs so many loops is that this
process can involve any number of soft scatterings off of other
particles in the plasma [the number of such scatterings is summed over
by Eq.~(\ref{eq:foo})].  The complicated ``cross-rungs'' in the upper diagram
are the result of opening up one of the lines in any one of the
self-energy contributions.
For more discussion of this point, see Ref.~\cite{GDMSEWM02}.

An interesting problem for the future is to understand the accuracy
of leading-order calculations of transport coefficients by calculating
higher order effects explicitly.
In the case of the QCD free energy it is known that,
beyond the ideal gas result, 
the perturbative expansion in powers of $g(T)$ is quite poorly behaved
\cite {hot-free1,hot-free2,hot-free3},
except for unrealistically small values of $g(T)$.
(Specifically, $T$ must substantially exceed the Planck scale.)
Does this same unpleasant behavior apply to transport coefficients?
At the moment, the only known test case is a many flavor limit
of QCD \cite {largeN},
where the leading-order result (as well as the next-to-leading-log
approximation thereto) is quite successful --- its accuracy is comparable
to the renormalization point sensitivity.
It would be useful to know if this holds more generally.

\fi

\begin{acknowledgments}

This work was supported, in part,
by the U.S. Department of Energy under Grant Nos.~DE-FG03-96ER40956
and DE-FG02-97ER41027.

\end{acknowledgments}


\appendix

\section {\boldmath $\twotwo$ Matrix Elements}\label{app:22}

\begin{figure}
\includegraphics[scale=0.40]{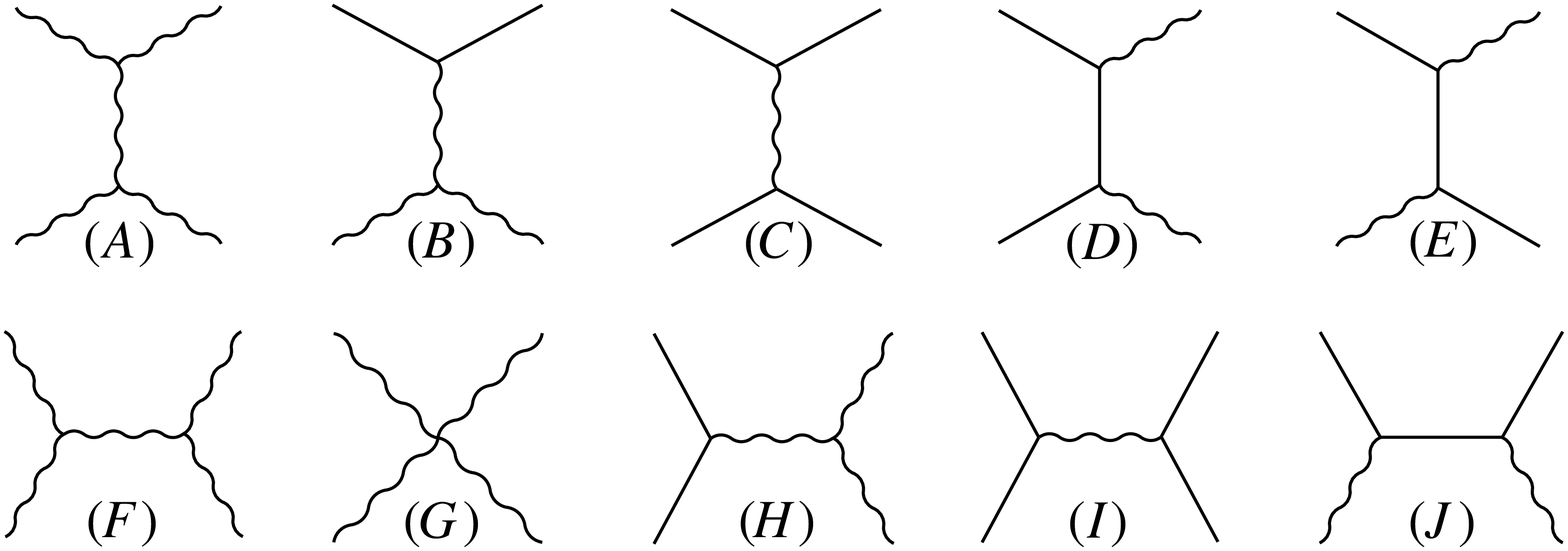}
\caption{\label{fig:diagrams} Diagrams for $2\leftrightarrow 2$ processes
needed at leading order in the coupling.  Leading-log calculations
require only the squares of diagrams $A$--$E$.
Next-to-leading-log,
or full leading order calculations
require evaluating the squares of diagrams $A$--$E$ with HTL self-energies
inserted on the internal lines, and then adding the (undressed) squares of
diagrams $F$--$J$, as well as the
interference terms between diagrams with the 
same initial and final states.}
\end{figure}

The matrix elements for all $2\lra2$ particle processes in a QCD-like
theory, neglecting thermal self-energy corrections,
are listed in Table \ref {table:mat_elements}.
These matrix elements
arise from the diagrams shown in Fig.~\ref{fig:diagrams}.
Terms in Table \ref {table:mat_elements}
with underlined denominators are sufficiently infrared sensitive
that thermal self-energy corrections must be included,
as discussed in Ref.~\cite{AMY5}.
Singly-underlined denominators indicate IR sensitive contributions
arising from soft gauge boson exchange,
while double-underlined denominators indicate
IR sensitive contributions from a soft exchanged fermion.

\begin{table}
\begin{center}
\begin{tabular}{|c|@{\quad}l@{\quad}|}
	\hline &
\\[-12pt]
	$ab \lra cd$ & $\qquad \left|{\cal M}^{ab}_{cd}\right|^2 / g^4$ 
\\[4pt]
	\hline &
\\[-12pt]
	$
	    \begin {array}{c}
		q_1 q_2 \lra q_1 q_2 \,,
		\\ q_1 \bar q_2 \lra q_1 \bar q_2 \,,
		\\ \bar q_1 q_2 \lra \bar q_1 q_2 \,,
		\\ \bar q_1 \bar q_2 \lra \bar q_1 \bar q_2
	    \end {array}
	$
    &
	$
	    \displaystyle
	    8\,  \frac{\df^2 \, \cf^2}{\da}
	    \left( \frac{s^2+u^2}{\ul{t^2}} \right)
	$
\\[28pt]
	$
	\begin {array}{c}
		q_1 q_1\lra q_1 q_1 \,, \\
		\bar q_1 \bar q_1 \lra \bar q_1 \bar q_1 \>
	\end{array}
	$
    & 
	$ \displaystyle
	    8\,  \frac{\df^2 \, \cf^2} {\da} 
	    \left( \frac{s^2+u^2}{\ul{t^2}} + \frac{s^2+t^2}{\ul{u^2}} \right)
	    +
	    16\,  \df \, \cf
	    \left( \cf {-} \frac{\ca}{2} \right) \frac{s^2}{tu}
	$
\\[14pt]
	$q_1 \bar q_1 \lra q_1 \bar q_1$
    &
	$ \displaystyle
	    8\,  \frac{\df^2 \, \cf^2}{\da} 
	    \left( \frac{s^2+u^2}{\ul{t^2}} + \frac{t^2+u^2}{s^2} \right)
	    +
	    16\, \df \, \cf
	    \left( \cf {-} \frac{\ca}{2} \right) \frac{u^2}{st}
	$
\\[14pt]
	$q_1 \bar q_1 \lra q_2 \bar q_2$
    &
	$ \displaystyle
	    8\, \frac{\df^2 \, \cf^2}{\da}
	    \left( \frac{t^2 + u^2}{s^2} \right)
	$
\\[10pt]
	$q_1 \bar q_1 \lra g \, g$
    &
	$ \displaystyle
	    8\, \df \, \cf^2
	    \left( \frac{u}{\ul{\ul{t}}} + \frac{t}{\ul{\ul{u}}} \right)
	    -
	    8\, \df \, \cf \, \ca
	    \left( \frac{t^2+u^2}{s^2} \right)
	$
\\[14pt]
	$
	\begin {array}{c}
	    q_1 \, g \lra q_1 \, g \,,\\ \bar q_1 \, g \lra \bar q_1 \, g
	\end {array}
	$
    &
	$ \displaystyle
	    -8\, \df \, \cf^2
	    \left( \frac{u}{s}  +  \frac{s}{\ul{\ul{u}}} \right)
	    +
	    8\, \df \, \cf \, \ca
	    \left( \frac{s^2 + u^2}{\ul{t^2}} \right)
	$
\\[14pt]
	$g \, g \lra g \, g$
    &
	$ \displaystyle
	    16\, \da \, \ca^2
	    \left(
		3 - \frac{su}{\ul{t^2}} - \frac{st}{\ul{u^2}} - \frac{tu}{s^2}
	    \right)
	$
\\[8pt]
\hline
\end{tabular}
\end{center}
\vspace*{-5pt}
\caption
    {%
    \label{table:mat_elements}
    Squares of
    vacuum matrix elements for $2\lra2$ particle processes
    in QCD-like theories, summed over spins and colors
    of all four particles.
    $q_1$ and $q_2$ represent fermions of distinct flavors,
    $\bar q_1$ and $\bar q_2$ are the associated antifermions,
    and $g$ represents a gauge boson.
    Note that the process $q_1 q_2 \lra q_1 q_2$, for example,
    appears $2\nf(\nf{-}1)$ times in the sum $\sum_{abcd}$
    over species in the linearized collision operator (2.19), while
    $q_1 \bar q_1 \lra q_1 \bar q_1$ and
    $q_1 \bar q_1 \lra gg$ each appear $4\nf$ times,
    $g g \lra g g$ appears just once, {\em etc.}
    }
\end{table}

\subsection{Self-energy corrections to matrix elements}
\label{app:mat_elements}

\subsubsection*{Fermion self-energy}

When $\twotwo$ particle processes involving $t$ (or $u$) channel
fermion exchange are computed using free propagators,
the resulting
squared matrix elements
(shown with double-underlined denominators in Table~\ref{table:mat_elements})
generate logarithmic infrared divergences in the collision term (\ref {eq:C22}).
This logarithmic infrared sensitivity is cut off by the inclusion
of the retarded thermal self-energy $\Sigma(Q)$, so that the internal
fermion propagator appearing inside the matrix element ${\cal M}$
is $[\slashchar Q - \slashchar \Sigma(Q)]^{-1}$.
Its conjugate, the advanced self energy $\Sigma^*(Q)$, appears in ${\cal M}^*$.
Since the exchange four-momentum $Q^\mu \equiv (\omega,\q)$ is spacelike,
and the thermal self-energy is only relevant when $Q$ is soft,
we only need the self-energy for spacelike momenta, 
$q\equiv |\q| > |\omega|$, in the hard thermal loop limit, $q \ll T$.
In this regime the self-energy was originally evaluated by Klimov
\cite {Klimov3} (and independently by Weldon \cite {Weldon2}), and is given by
\begin{eqnarray}
\Sigma^0(Q) & = & \frac{\mQ^2}{2q} \left[
	\ln \left(\frac{q+\omega}{q-\omega}\right) - i \pi \right]
	\, ,
\\
{\bSigma}(Q) & = & -\hat\q \, \frac{\mQ^2}{q}
	\left\{ 1 - \frac{\omega}{2q} \left[
	\ln \left(\frac{q+\omega}{q-\omega}\right) - i \pi \right] \right\}
	\, .
\end{eqnarray}
Here $\mQ = \sqrt{ \cf /8} \> gT$ is the
(leading-order) ``fermion thermal mass,''
equal to the thermal energy of a fermion at zero momentum.  

For the process $fg \to fg$,
the net effect of the inclusion of the fermion self-energy is to make
the replacement
\begin {eqnarray}
    \frac {s}{u}
    &\longrightarrow&
    \frac {
	-{\textstyle \frac 18} \, {\rm tr}
	\left[
	\slashchar P \, \gamma^\mu \, \slashchar {{\cal Q}} \, \gamma^\nu
	\slashchar P' \, \gamma_\nu \, \slashchar {{\cal Q}}^* \, \gamma_\mu \,
        \right]
	  }{\left| {\cal Q} \cdot {\cal Q} \right|^2}
	=
	\frac {
	    -4 \, \Re [ (P\cdot {\cal Q}) (P'\cdot {\cal Q}^*) ]
	    + t \, {\cal Q} \cdot {\cal Q}^*
	}{\left| {\cal Q} \cdot {\cal Q} \right|^2} \,,
\end {eqnarray}
[with $(-+++)$ metric convention]
in the contribution shown in Table \ref {table:mat_elements}.
Here $P^\mu = (|\p|,\p)$ and $P'^\mu = (|\p'|,\p')$ are the incoming and
outgoing fermion 4-momenta, respectively, and
${\cal Q}^\mu \equiv P^\mu - K'^\mu - \Sigma^\mu(P{-}K')$
with $K'$ the outgoing gauge boson 4-momentum.

The analogous replacements needed
in the $ff \lra gg$ squared matrix elements are
\begin {eqnarray}
    \frac {u}{t}
    &\longrightarrow&
	\left.
	\frac {
	    4 \, \Re [ (P\cdot {\cal Q}) (K\cdot {\cal Q}^*) ]
	    + s \, {\cal Q} \cdot {\cal Q}^*
	}{\left| {\cal Q} \cdot {\cal Q} \right|^2}
	\right|_{{\cal Q}^\mu = P^\mu - P'^\mu - \Sigma^\mu(P{-}P')}
	\,,
\\
\noalign{\hbox{and}}
    \frac {t}{u}
    &\longrightarrow&
	\left.
	\frac {
	    4 \, \Re [ (P\cdot {\cal Q}) (K\cdot {\cal Q}^*) ]
	    + s \, {\cal Q} \cdot {\cal Q}^*
	}{\left| {\cal Q} \cdot {\cal Q} \right|^2}
	\right|_{{\cal Q}^\mu = P^\mu - K'^\mu - \Sigma^\mu(P{-}K')}
	\,,
\end {eqnarray}
where $P$ and $K$ are the incoming fermion momenta,
and $P'$ and $K'$ the outgoing gauge boson momenta.

\subsubsection*{Gauge boson self-energy}

Processes involving $t$ or $u$ channel gauge boson exchange
require inclusion of the thermal gauge boson self-energy
on the internal propagator
to cut off the infrared sensitivity of these processes.
Because the self-energy only matters when the exchange momentum is soft,
one may exploit the fact that soft gluon exchange between hard particles
is spin-independent (to leading order) \cite {AMY4}.
If one separates the IR sensitive matrix elements (those with
singly-underlined denominators in Table \ref {table:mat_elements})
into (i) the result one would have with fictitious scalar quarks
plus (ii) a spin-dependent remainder, then all the IR sensitivity
resides in the first spin-independent piece.
This is the only piece which must be recomputed with the thermal
self-energy included.
For the $t$-channel exchange terms, this amounts to using the exact 
identities
\begin {equation}
    \frac{s^2{+}u^2}{t^2} = \half + \half \, \frac{(s{-}u)^2}{t^2} \,, \qquad
    \frac{su}{t^2} = \quarter - \quarter \, \frac{(s{-}u)^2}{t^2} \,,
\end {equation}
and then replacing
\begin {equation}
    \frac {(s{-}u)^2}{t^2} \longrightarrow
    \left|
	D(P{-}P')_{\mu\nu}(P{+}P')^\mu (K{+}K')^\nu
    \right|^2 ,
\label {eq:sut}
\end {equation}
where $D(Q)_{\mu\nu}$ is the retarded thermal equilibrium
gauge field propagator, evaluated in the HTL approximation.

The HTL result for the above replacement (\ref{eq:sut}) does not
depend on gauge choice.  One convenient choice is Coulomb gauge,
where \cite{BraatenPisarski}
\begin{eqnarray}
    D_{00}(\omega,\q) & = &
    \frac{-1}{q^2 + \Pi_{00}(\omega,q)} \, ,
\\[3pt]
    D_{ij}(\omega,\q) & = & \frac{\delta_{ij} - \hat\q_i \, \hat\q_j}
	    {q^2 - \omega^2 + \PiT(\omega,q)} \, ,
\\[6pt]
    D_{0i}(\omega,\q) & = & D_{i0}(\omega,\q) \: = \: 0 \, .
\end{eqnarray}
The equilibrium transverse and longitudinal gauge boson self-energies
are \cite{Klimov1,Weldon1}
\begin {eqnarray}
    \PiT(\omega,\q) & = & \mD^2
	\left\{
	    \frac{\omega^2}{2q^2} 
	+
	    \frac{\omega \, ( q^2 {-} \omega^2 )}{4 q^3}
	    \left[
		\ln \left( \frac{q+\omega}{q-\omega}\right) - i \, \pi
	    \right]
	\right\}
	\, , 
\label{eq:PiT} \\
    \Pi_{00}(\omega,\q) & = & \mD^2
	\left\{
	    1
	    -
	    \frac{\omega}{2q}
	    \left[
		\ln \left(\frac{q+\omega}{q-\omega}\right) - i \, \pi
	    \right]
	\right\} \, ,
\label{eq:PiL}
\end {eqnarray}
where we have assumed $|\omega| < q$,
which is the only case of relevance.

\subsection {Integration variables}

Since all external particles are to be treated as massless,
the domain of the phase space integrations appearing in
matrix elements of the linearized collision operator (\ref{eq:C22})
are the same for all $\twotwo$ processes.
One rather straightforward method for doing the multi-dimensional
numerical integration is to use an adaptive Monte Carlo integrator.
This can give reasonable accuracy at a tolerable investment of
computational effort, but for the highest accuracy it is preferable
to use nested one-dimensional adaptive Gaussian integration.

In order to handle efficiently the infrared-sensitive terms
in $t$ and $u$ channel processes which give rise to leading-log contributions,
it is useful to pick the exchange momentum $q$ and energy $\omega$
as two of the integration variables.
When doing nested adaptive quadrature integrations,
it is especially advantageous to choose integration variables
in a manner which allows one to perform analytically
as many of the integrations as possible.
In particular, it is convenient to use different parameterizations
for terms in Table \ref{table:mat_elements} having different denominators.
For terms having denominators of $t$ (or $t^2$), the $t$ channel
parameterization described below allows all but four integrations
to be done analytically.
And similarly, the $u$ and $s$ channel parameterizations described below
allow all but four integrations to be done analytically
for terms with denominators of $u$ (or $u^2$), or $s$ (or $s^2$),
respectively.
The constant term in the $gg \lra gg$ matrix element can be handled
using any of these parameterizations.
The only other terms in Table \ref{table:mat_elements}
are those involving $s^2/tu$ or $u^2/st$.
These can be reduced to the previous cases by rewriting
\begin {eqnarray}
   \frac{s^2}{tu} &=& - \frac{s}{t} - \frac{s}{u} \, ,
\qquad
   \frac{u^2}{st} = - \frac{u}{s} - \frac{u}{t} \, .
\label {eq:s2tu}
\end {eqnarray}
In what follows, our convention for labeling momenta in $2\lra2$
processes is that $P, K \lra P', K'$.


\subsubsection*{$t$ channel parameterization}\label{t_phase}

For terms containing
$t = -(P'-P)^2$ in the denominator,
it is convenient to use the spatial delta function in (\ref {eq:C22})
to perform the
$\k'$ integration, and to shift the $\p'$ integration into an
integration over $\p'{-}\p \equiv \q$.
The angular integrals may be written
in spherical coordinates defined such that
the $z$ axis is in the direction of $\q$
while $\p$ lies in the $xz$ plane.
This yields
\begin{eqnarray}
    \Big( \chi_\ij, \C^\twotwo \chi_\ij \Big)
    &=&
    \frac{\beta^3}{(4\pi)^6}
    \sum_{abcd}
    \int_{0}^{\infty} q^2 dq \> p^2 dp \> k^2 dk 
	\int_{-1}^{1} d \cos\theta_{pq} \> 
	d\cos\theta_{kq}
	\int_0^{2\pi} d\phi \;
	\frac{1}{p\,k\,p'\,k'}
\nonumber \\ && \hspace {0.9cm} {}\times
	\left|{\cal M}^{ab}_{cd}\right|^2 \>
	\delta(p{+}k{-}p' {-} k') \>
	f^a_0(p) \, f^b_0(k) \, [1{\pm}f^c_0(p')] \, [1{\pm}f^d_0(k')]
\nonumber \\ && \hspace {0.9cm} {}\times
	\left[
	    \chi^a_\ij(\p) + \chi^b_\ij(\k) - \chi^c_\ij(\p') - \chi^d_\ij(\k')
	\right]^2
	 \, ,
\end{eqnarray}
where $p$, $k$, and $q$ denote to the magnitudes of the
corresponding three-momenta (not the associated 4-momenta),
$p' \equiv |\q+\p|$ and $k' \equiv |\k-\q|$ are the magnitudes of the 
outgoing momenta,
$\phi$ is the azimuthal angle of $\k$ (and $\k'$)
[{\em i.e.}, the angle between the $\p$,$\q$ plane and the $\k$,$\q$ plane],
and $\theta_{pq}$ is the angle between $\p$ and $\q$
(so $\cos\theta_{pq} \equiv \hat\p \cdot \hat\q$), {\em etc}.

Following Baym {\it et~al.}~\cite{BMPRa}, it is convenient to
introduce a dummy
integration variable $\omega$, defined to equal
the energy transfer $p' - p$,
so that
\begin{equation}
    \delta(p+k-p' - k')
    =
    \int_{-\infty}^\infty d\omega \;
    \delta(\omega + p - p') \> \delta(\omega-k+k') \, .
\end{equation}
Evaluating $p'=|\p+\q|$
in terms of $p$, $q$, and $\cos \theta_{pq}$, and 
defining $t = \omega^2 - q^2$ (which is the usual Mandelstam variable),
one finds
\begin{eqnarray}
    \delta(\omega+p-p')
    &=&
    \frac{p'}{pq} \>
    \delta\biggl( \cos \theta_{pq} - \frac{\omega}{q} 
	- \frac{t}{2pq} \biggr) \> \Theta(\omega + p )\,,
\\
    \delta(\omega-k+k')
    &=&
    \frac{k'}{kq} \>
    \delta\biggl( \cos \theta_{kq} - \frac{\omega}{q} 
	+ \frac{t}{2kq} \biggr) \> \Theta ( k - \omega ) \,,
\end{eqnarray}
with $\Theta$ the unit step function.
The $\cos\theta$ integrals may now be trivially performed and yield one
provided $p > \half(q-\omega)$, $k>\half(q+\omega)$, and
$|\omega|<q$; otherwise the delta functions cannot both be satisfied.
The remaining integrals are
\begin{eqnarray}
    \Big( \chi_\ij, \C^\twotwo \chi_\ij \Big)
    &=&
    \frac{\beta^3}{(4\pi)^6}
    \sum_{abcd}
    \int_0^{\infty} dq
    \int_{-q}^q d\omega 
    \int_{\frac{q-\omega}{2}}^\infty dp 
    \int_{\frac{q+\omega}{2}}^\infty dk
    \int_0^{2\pi} d\phi
\nonumber \\ && \hspace {1cm} {}\times
	\left|{\cal M}^{ab}_{cd}\right|^2 \>
	f^a_0(p) \, f^b_0(k) \, [1 \pm f^c_0(p')] \, [1 \pm f^d_0(k')]
\nonumber \\ && \hspace {1cm} {}\times
	\left[
	    \chi^a_\ij(\p) + \chi^b_\ij(\k) - \chi^c_\ij(\p') - \chi^d_\ij(\k')
	\right]^2
	 \, ,
\la {eq:Q3}
\end{eqnarray}
with $p'=p+\omega$ and $k'=k-\omega$.
For evaluating the final factor of (\ref {eq:Q3}),
we use the relationship (\ref{eq:II}) that
\begin{equation}
    I_\ij(\hat\p) \, I_\ij(\hat\k)
    =
	P_\ell(\cos\theta_{pk}) \,.
\la{eq:contract_I}
\end{equation}
One therefore needs expressions for the angles between all species,
as well as the remaining Mandelstam variables $s$ and $u$, which may 
appear in ${\cal M}^2$.
They are
\begin {subequations}
\begin {eqnarray}
\la{eq:s_equals}
    s &=& \frac{-t}{2q^2}
	\left\{
	\left[(p+p')(k+k')+q^2 \right]
	    - \cos\phi \>
		\sqrt{\textstyle
		\left(4pp'+t \right)\left(4kk'+t \right) } \,
	\right\}
	\, , \\[3pt]
u & = & -t -s \, ,
\label{eq:u_equals}
\end{eqnarray}
\end {subequations}
and
\begin {subequations}
\begin{eqnarray}
\cos \theta_{pq}\: & = & \frac{\omega}{q} + \frac{t}{2pq} 
	\, , \hspace{1.07in} \;
\cos \theta_{p'q} \: = \frac{\omega}{q} - \frac{t}{2p'q} \, , \\
\cos \theta_{kq} \: & = & \frac{\omega}{q} - \frac{t}{2kq} 
	\, , \hspace{1.1in}
\cos \theta_{k'q}\: = \frac{\omega}{q} + \frac{t}{2k'q} \, ,  \\
\cos \theta_{pp'} & = & 1 + \frac{t}{2pp'} \, ,\hspace{1.12in}
\label{eq:with_t}
\cos \theta_{kk'} \: = 1 + \frac{t}{2kk'} \, , \\
\cos \theta_{pk'} & = & 1 + \frac{u}{2pk'}
	\, , \hspace{1.115in}
\label{eq:with_u}
\cos \theta_{p'k} \: = 1 + \frac{u}{2p'k}
	\, , \\
\cos \theta_{pk} \: & = & 1 - \frac{s}{2pk}
	\, , \hspace{1.15in}
\label{eq:with_s}
\cos \theta_{p'k'} = 1 - \frac{s}{2p'k'}
	\, .
\label{eq:angles}
\end {eqnarray}
\end {subequations}
For terms in which
only $t$ appears in the denominator of the matrix element, the
$\phi$ integration can be easily performed analytically, leaving four
integrals which must be evaluated numerically.

\subsubsection*{$u$ channel parameterization}

For terms in which $u=-(K'-P)^2$ appears in the denominator,
exchanging $\p'$ and $\k'$ in the $t$ channel parameterization
provides the natural choice of variables.

\subsubsection*{$s$ channel parameterization}

For terms in which $s = -(P+K)^2$ appears in the denominator,
one may use the spatial delta function in Eq.~(\ref{eq:C22}) to perform the
$\k'$ integration, and then shift
the $\k$ integration to an integral over $\q = \p + \k$,
the total incoming spatial momentum
(and the momentum on the internal propagator in $s$ channel exchange processes).
Again choosing spherical coordinates so that $\q$ lies on the $z$ axis and $\p$
lies in the $xz$ plane,
the $\twotwo$ contribution (\ref{eq:C22}) becomes
\begin{eqnarray}
    \Big( \chi_\ij, \C^\twotwo \chi_\ij \Big)
    &=&
    \frac{\beta^3}{(4\pi)^6}
    \sum_{abcd}
    \int_{0}^{\infty} q^2 dq \> p^2 dp \> {p'}^2 dp' 
	\int_{-1}^{1} d \cos\theta_{pq} \> 
	d\cos\theta_{p'q}
	\int_0^{2\pi} d\phi \>
	\frac{1}{p\,k\,p'\,k'}
\nonumber \\ && \hspace {0.9cm} {}\times
	\left|{\cal M}^{ab}_{cd}\right|^2 \>
	\delta(p{+}k{-}p' {-} k') \>
	f^a_0(p) \, f^b_0(k) \, [1 \pm f^c_0(p')] \, [1 \pm f^d_0(k')]
\nonumber \\ && \hspace {0.9cm} {}\times
	\left[
	    \chi^a_\ij(\p) + \chi^b_\ij(\k) - \chi^c_\ij(\p') - \chi^d_\ij(\k')
	\right]^2
	 \, ,
\end{eqnarray}
where now $k=|\q-\p|$, $k'=|\q-\p'|$, and $\phi$ is the azimuthal angle
of $\k$ (and $\k'$).
Introducing the total energy $\omega$ via
\begin{equation}
\delta(p+k-p'-k') = \int_{0}^{\infty} d\omega \; \delta(\omega - p - k)
	\> \delta ( \omega - p' - k' ) \, ,
\end{equation}
and defining $s = \omega^2 - q^2$
(which is the usual Mandelstam variable), one finds
\begin{eqnarray}
\delta(\omega - p - k) & = & \frac{k}{pq} \: \, \delta \biggl( 
	\cos \theta_{pq} \: - \frac{\omega}{q} 
	+ \frac{s}{2pq} \: \biggr) \> \Theta(\omega - p \:) \, , \\
\delta(\omega - p' - k') & = & \frac{k'}{p'q} \, \delta \biggl( 
	\cos \theta_{p'q} - \frac{\omega}{q} 
	+ \frac{s}{2p'q} \biggr) \> \Theta(\omega - p') \, .
\end{eqnarray}
Integration over $\cos \theta_{pq}$ and $\cos \theta_{p'q}$ yields unity
provided $q < \omega$, $|2p - \omega| < q$, and $|2 p' - \omega| < q$
(and zero otherwise).
Therefore,
\begin{eqnarray}
\Big( \chi_\ij, \C^\twotwo \chi_\ij \Big)
	&=&
	\frac{\beta^3}{(4\pi)^6} \sum_{abcd} \>
	\int_0^\infty \! d\omega \int_0^\omega \! dq 
	\int_{\frac{\omega-q}{2}}^{\frac{\omega+q}{2}} \! \! dp
	\int_{\frac{\omega-q}{2}}^{\frac{\omega+q}{2}} \! \! dp'
	\int_0^{2\pi} \! d\phi  \nonumber \\
	&& \hspace{1cm} \times \left| {\cal M}^{ab}_{cd}\right|^2 \>
	f_0^a(p) \, f_0^b(k) \, [1{\pm}f_0^c(p')] \,
	[1{\pm}f_0^d(k')] \nonumber \\
	&& \hspace{1cm} \times 
	\left[
	    \chi^a_\ij(\p) + \chi^b_\ij(\k) - \chi^c_\ij(\p') - \chi^d_\ij(\k')
	\right]^2
	 \, ,
\end{eqnarray}
with $k=\omega-p$ and $k'=\omega-p'$.  The other Mandelstam variables are
\begin {subequations}
\begin{eqnarray}
t & = & \frac{s}{2q^2} \left\{ \left[ (p-k)(p'-k') - q^2 \right] 
	+ \cos \phi \sqrt{(4pk-s)(4p'k'-s)} 
	\right\} \, , \\
u & = & -s -t \, ,
\end{eqnarray}
\end {subequations}
and the angles between $\q$ and the external momenta are 
\begin {subequations}
\begin{eqnarray}
\cos \theta_{pq} & = & \frac{\omega}q - \frac{s}{2pq}
	\, , \hspace{1.43in}
\cos \theta_{p'q} = \frac{\omega}q - \frac{s}{2p'q}
	\, , \\
\cos \theta_{kq} & = & \frac{\omega}q - \frac{s}{2kq}
	\, , \hspace{1.42in}
\cos \theta_{k'q} = \frac{\omega}q - \frac{s}{2k'q}
	\, .
\end{eqnarray}
\end {subequations}
Eqs.~(\ref{eq:with_t})--(\ref{eq:with_s}) for the angles between
external momenta still hold.
For terms in which only $s$ appears in a denominator,
the $\phi$ integration can be easily performed analytically,
leaving four integrals to do numerically.

\section{Matrix elements for $\onetwo$ processes}
\label{app:12}

In this appendix, we will review the integral equations that determine
the splitting/joining functions $\gamma^a_{bc}$ for effective
$1 \lra 2$ processes,
appearing in our various formulations
(\ref{eq:collision12}) and (\ref{eq:C12}) of the Boltzmann collision term.
These equations are summarized in Ref.\ \cite{AMY5}, but here we will
make a number of simplifications applicable to the problem at hand.

The only $\onetwo$ processes with $O(g^4 T^4)$ total rates
per unit volume involve
hard, collinear creation or destruction of a gauge boson.
In QCD, the relevant processes are $q \lra qg$, $\bar q \lra \bar q g$,
$g \lra q \bar q$, and $g \lra g g$.
By ``hard'' we mean that
the outgoing states each have $O(T)$ energy, and by ``collinear'' we
mean that the angles between the three external momentum vectors are all
$O(g)$.  [These parametrically small opening angles are ignored when
evaluating distribution functions in the $1 \lra 2$
piece of our collision term (\ref{eq:collision12}), which only makes
a sub-leading error in the evaluation of transport coefficients.]

The duration of such processes (also known as the formation time of the gauge
boson) is parametrically $O(1/g^2 T)$, which is the same as the mean free path
for hard particles to undergo soft scattering (with momentum exchange of order
$gT$).  For this reason, one must sum all possible number of soft scatterings
with other excitations during the emission process.  (In realistic
theories, at least one such soft scattering is required to allow for
energy-momentum conservation in the $1\lra2$ process.)
This summation can be implemented by an integral equation which must be solved
numerically.
For a complete discussion, and a derivation of the initial
integral equation presented below, see Refs.~\cite{AMY5,AMY2,AMY4}.

\subsection{The integral equation for \boldmath$\gamma^a_{bc}$}

For non-Abelian gauge theories such as QCD, the splitting/joining functions
$\gamma^a_{bc}(p';p,k)$ for particle types $a \lra bc$ with momenta
$p' \lra p k$ are given in equilibrium by \cite{AMY5}
\begin {subequations}
\label{eq:dgamma}
\begin {eqnarray}
	\gamma^q_{qg}(p'; p, k)
   &=&
	\gamma^{\bar q}_{\bar qg}(p'; p, k)
   =
	\frac{{\smash{p'}\vphantom p}^2 + p^2}
             {{\smash{p'}\vphantom p}^2 \, p^2 \, k^3} \>
	{\cal F}_{\rm q}(p',p,k) \,,
\\
	\gamma^g_{q\bar q}(p'; p, k)
   &=&
	\frac{p^2 + k^2}
             {k^2 \, p^2 \, {\smash{p'}\vphantom p}^3} \>
	{\cal F}_{\rm q}(k,-p,p') \,,
\\
	\gamma^g_{gg}(p'; p, k)
   &=&
	\frac{{\smash{p'}\vphantom p}^4 + p^4 + k^4}
             {{\smash{p'}\vphantom p}^3 \, p^3 \, k^3} \>
	{\cal F}_{\rm g}(p',p,k) \,,
\end {eqnarray}
\end {subequations}
where
\begin {equation}
   {\cal F}_s(p',p,k) \equiv
     \frac{d_s \, C_s \, \alpha}{2(2\pi)^3}\,
     \int \frac{d^2 h}{(2\pi)^2} \; 2\h \cdot \Re\, \F_s(\h;p',p,k)
\label {eq:eoo}
\end {equation}
and $\alpha \equiv g^2/(4\pi)$.
The function $\F_s(\h;p',p,k)$,
for fixed given values of $p'$, $p$, and $k$,
depends on a two-dimensional vector $\h$ which is related to
transverse momentum during the splitting process.
$\F_s$ is the solution to the linear integral equation
\begin {align}
    2 \h
    =
    i \, \delta E(\h;p',p,k) \, \F_s(\h;p',p,k)
\hspace {2em} &
\nonumber\\
        + g^2 \int \frac{d^2 \qt}{(2\pi)^2} \,
            {\cal A}(\qt) \,
	\biggl\{
	     (C_s - \half \ca) & \left[{\F}_s(\h;p',p,k)
                       - {\F}_s(\h{-}k\,\qt;p',p,k) \right]
\nonumber\\
	     + \half \ca & \left[{\F}_s(\h;p',p,k)
                       - {\F}_s(\h{+}p'\qt;p',p,k) \right]
\nonumber\\
	     + \half \ca & \left[{\F}_s(\h;p',p,k)
                       - {\F}_s(\h{-}p\,\qt;p',p,k) \right]
	\biggr\} 
	\,,
\label {eq:foo}
\end {align}
where
\begin {equation}
    \delta E(\h;p',p,k)
    =
	\frac{\mg^2}{2k} + \frac{\meffs^2}{2p} - \frac{\meffs^2}{2p'} 
	+ \frac{\h^2}{2p \, k\, p'}
\label {eq:deltaE}
\end {equation}
and
\begin {equation}
   {\cal A}(\qt) \equiv
       \int \frac{dq^z}{2\pi}
	    \Dlangle\strut A^-(Q) [A^-(Q)]^* \Drangle \biggl|_{q^0=q^z} .
\label {eq:A}
\end {equation}
Here $Q = (q^0,\qt,q^z)$, $A^- \equiv A^0 - A^z$,
and $\Dlangle\strut A^-(Q) [A^-(Q)]^* \Drangle$
is the thermal Wightman correlator
evaluated in the hard-thermal-loop approximation.
Explicit formulas for this correlator in equilibrium may be found, for example,
in Ref.\ \cite{AMY4}.
However, we will make use here of the wonderful simplification found by
Aurenche {\em et al.}\ \cite{Gelis2} showing that the integral
(\ref {eq:A}) has a remarkably simple form,
\begin {equation}
   {\cal A}(\qt) = T\left(\frac{1}{|\qt|^2} - \frac{1}{|\qt|^2+\mD^2}\right).
\end {equation}

\subsection {Variational solution}

One way to solve the integral equation for $\F_s$ is to use a variational
method, similar to the method used in the main text for the Boltzmann equation
and implemented in the Abelian case for $1 \lra 2$ processes in
Ref.\ \cite{AMY3}.  Some further simplifications are then possible, as we
shall describe below.
The variational formulation is
\begin {equation}
   {\cal F}_s(p',p,k) \equiv
     \frac{d_s \, C_s \, \alpha}{(2\pi)^3}\,
     (\QLPM_s)_{\rm extremum} ,
\end {equation}
with
\begin {equation}
   \QLPM_s[\F] =
   \Re\left[
      \Bigl(2\h, \F\Bigr)
       - \half \Bigl(\F, (i\,\delta E + \KLPM_s) \F\Bigr)
   \right] ,
\end {equation}
\begin {equation}
   (\QLPM_s)_{\rm extremum} =
   {\textstyle\frac12} \, \Re
   \Bigl( 2\h, (i \, \delta E + \KLPM_s)^{-1} 2\h \Bigr) ,
\end {equation}
where in this context
\begin {equation}
   ({\bm f},{\bm g}) \equiv \int \frac{d^2\h}{(2\pi)^2}
   \, {\bm f(\h)} \cdot {\bm  g}(\h) .
\end {equation}
The analog $\KLPM_s$ of the collision operator is given by
\begin {equation}
   \Bigl( \F, \KLPM_s \F \Bigr) =
   g^2 \Bigl\{
     (C_s {-} \half \ca) \, \Bigl(\F, K(k)\, \F\Bigr)
            + \half\ca \, \Bigl(\F, K(-p')\, \F\Bigr)
            + \half\ca \, \Bigl(\F, K(p)\, \F\Bigr)
   \Bigr\} ,
\label {eq:KLPMdef}
\end {equation}
where
\begin {equation}
   \Bigl(\F, K(\mom)\, \F\Bigr) \equiv 
     \half \int \frac{d^2\h}{(2\pi)^2} \, \frac{d^2\qt}{(2\pi)^2} \,
     {\cal A}(\qt) \, \Bigl[ \F(\h) - \F(\h-\mom\qt) \Bigr]^2 .
\label {eq:Kdef}
\end {equation}
Rotational invariance implies that the extremum must have the
rotationally covariant form
\begin {equation}
   {\F}_s(\h;p',p,k) = \h \, \FF_s(h;p',p,k) .
\end {equation}
The number of integrations necessary to evaluate
$\bigl(\F, K(\mom)\, \F\bigr)$ can then be
reduced by 
expanding the square in (\ref{eq:Kdef}), switching integration variables
to the dimensionless variables ${\bm u} = \h/(\mom\mD)$ and
${\bm w} = (\h-\mom\qt)/(\mom\mD)$,
and then performing the angular
integrations using the explicit form (\ref{eq:A}) for ${\cal A}(\qt)$:
\begin {align}
   (\h \FF, K(\mom) \, \h \FF) =
   T\, \frac{(\mom\mD)^4}{(4\pi)^2} \int_0^\infty du \> u \> dw \> w \>
&
   \left\{
     \frac{u^2{+}w^2}{|u^2{-}w^2|}
     -
     \frac{(u^2+w^2+1)}{
         \bigl[\bigl(\bigl(u{+}w)^2+1\bigr)\bigl((u{-}w)^2+1\bigr)\bigr]^{1/2}
     }
   \right\}
\nonumber\\ & \times
   \bigl[\FF(\mom\mD u) - \FF(\mom\mD w)\bigr]^2 .
\label {eq:plup}
\end {align}

Making a finite basis expansion of the form,
\begin {equation}
   \FF(h) = \sum_{m=1}^{N_{\rm r}} \FF_{\rm r}^{(m)} \, \Phi_{\rm r}^{(m)}(h)
        + i \sum_{m=1}^{N_{\rm i}} \FF_{\rm i}^{(m)} \, \Phi_{\rm i}^{(m)}(h)
   ,
\end {equation}
an efficient set of basis elements is \cite{AMY3}
\begin{eqnarray}
\Phi^{(m)}_{\rm r}(h) &=& \frac{ 
	(h^2/A)^{m-1}}
	{\left( 1+h^2/A \right)^{N_{\rm r}+2}} \, ,
	\qquad m = 1, ..., N_{\rm r} \,,
\\
\Phi^{(m)}_{\rm i}(h) &=& \frac{ 
	(h^2/A)^{m-1}}
	{\left( 1+h^2/A \right)^{N_{\rm i}}} \, ,
	\qquad \quad m = 1, ..., N_{\rm i} \,,
\end{eqnarray}
where the scale $A$ is chosen to optimize convergence as the basis
set is increased.%
\footnote{
   An efficient choice of $A$ can be found by numerical search as described
   in Ref.\ \cite{AMY3}.
   However, in the present context, we have also found that the simple choice
   $A = p p' \mg^2 + k (p'{-}p) \meffs^2$ works reasonably well.
   This is the value of $h^2$ for which the last term in Eq.\
   (\ref{eq:deltaE}) equals the preceding terms.
}
Since the basis elements $\Phi$ are only functions of $h^2/A$, one can
see from (\ref{eq:plup}) that matrix elements such as
$(\Phi_{\rm r}^{(m)}, K(\mom) \, \Phi_{\rm r}^{(n)})$
will equal $T(\mD \mom)^4$ times a dimensionless function depending
only on $m$, $n$, and
$z \equiv (\mD \mom)^2/A$.  That means that one can numerically
generate these functions just once for a given basis size (by evaluating
on a fine mesh of points in $z$ and spline interpolating), and
then repeatedly re-use their tabulated values for calculations
in different theories with different values of $\mD$ and $A$.

For further details relating to setting up the framework for numerical
evaluation, see the treatment of photo-emission in Ref.\ \cite{AMY3}.
The solution of the integral equation
is somewhat involved but can be performed with very
good numerical accuracy ($<10^{-4}$ relative errors).  Once the values
of the $K(\mom)$ matrix elements have been tabulated, the solution
of the splitting functions $\gamma^a_{bc}$ is quite fast.
Consequently,
even though these splitting functions appear inside the integrals
of the Boltzmann collision term (\ref{eq:C12}), the
numerical cost of nested integral equations
is ultimately not very large.

\section{Asymptotic behavior of inverse log expansion}
\label{app:converge}

Here we prove that the expansion (\ref{eq:Qseries})
in inverse powers of $\ln (\mu/\mD)$
is an asymptotic series with vanishing radius of convergence.
We also give a (non-rigorous) argument that the
expansion coefficients have non-alternating factorial growth,
implying the presence of singularities in the Borel transform
on the positive real axis.

\subsection{What we need to show}

Within the $\ell = 2$ or C-odd $\ell = 1$ symmetry channels of interest,
both the linearized collision operator $\C$,
and its leading-log piece $\A$ defined in Eq.~(\ref{eq:partialg}),
are real, symmetric, positive definite linear operators.%
\footnote
    {
    More generally, $\C$ is positive semi-definite with zero modes
    associated with conserved charges: one
    C-even $\ell = 0$ zero mode associated with energy conservation,
    one C-even $\ell = 1$ zero mode associated with momentum conservation,
    and various C-odd $\ell=0$ zero modes associated with conservation of
    quark flavors.
    The leading-log operator $\A$ has the same structure except for
    one additional $\ell = 0$ zero mode
    associated with total particle number (gluon plus quark plus anti-quark)
    conservation.
    }
This can be seen
by looking at the explicit forms (\ref {eq:Q2}) and (\ref{eq:C_LL}).
In other words, within the symmetry channels of interest,
the operators $\A$ and $\C$ are both invertible.

However, we are not assured that the difference operator
$\dCbar (\mu/T)$
which generates the inverse log expansion is positive definite,
because
its definition (\ref{eq:lim1}) involves a limiting procedure in which
we subtract larger and larger multiples of $\A$ as we take the
$\mD/T \rightarrow 0$ limit.
There is no guarantee that adding
a finite multiple of $\A$ to $\dCbar(\mu/T)$ will yield a
positive operator, and in fact it does not.
In other words, even though $\C$ is a positive definite operator,
removing all subleading (in $g$) contributions by defining
\begin {equation}
    \C \big|_{\rm pure \; leading \; order}
    \equiv
    \CLL(\mu) + \delta \C(\mu)
    =
    g^4 T \left[ \ln (\mu/\mD) \> \A + \dCbar(\mu/T) \right] ,
\end {equation}
yields an operator which is not positive definite.
To show this, we will
demonstrate that there is a family $\phi^{(m)}$ of functions such that
\begin{equation}
\lim_{m \rightarrow \infty} \frac
	{\Big( \phi^{(m)}_\ij\, , \, \dCbar(\mu/T) \, \phi^{(m)}_\ij \Big)}
	{\Big( \phi^{(m)}_\ij \, , \, \A \, \phi^{(m)}_\ij \Big)} 
	= - \infty \, ,
\label{eq:bad}
\end{equation}
which is enough to ensure that
$[\ln(\mu/\mD) \, \A  + \dCbar(\mu/T)]$ cannot be
free of negative eigenvalues for any finite $\mu/\mD$.
Since $\A$ has positive spectrum, this implies that the expansion
(\ref{eq:C_inv}) in inverse powers of $\ln (\mu/\mD)$
must have a vanishing radius of convergence.%
\footnote{
    Here's a general proof.
    Let $C(t) = A + t B$, with $A$ an Hermitian positive definite
    and hence invertible operator, $B$ Hermitian, and $t$ real.
    A necessary (but not sufficient) condition for the convergence
    of the Taylor expansion of $C(t)^{-1}$ in powers of $t$ at $t = \bar t$,
    is the {\it existence}\/ of $C(t)^{-1}$ for all $t$ between 0 and
    $\bar t$.
    As $t$ increases from zero, $C(t)$ first fails to be invertible
    when some eigenvalue crosses zero.
    If $(v,C(\bar t) \, v) < 0$ for some vector $v$,
    then one or more eigenvalues
    must have crossed zero for $t < \bar t$, implying that $\bar t$
    is outside the radius of convergence.
}

\subsection{The test function sequence \boldmath$\{ \phi^{(m)} \}$}
\label {sec:phim}

In the limit $\mD/T\to0$, which is used to define $\A$ and
$\dCbar(\mu/T)$, the leading-log contribution dominates
transport coefficients.
The leading-log result is associated with
$t$-channel (or $u$-channel) $2\lra2$ processes with
momentum transfers $\q$ in the range $\mD \ll q \ll T$.
For the hard particle
momenta $\p \sim \k \sim T$ that dominate transport, one may then
approximate
\begin {subequations}
\label {eq:smallq_approx}
\begin {equation}
  \chi_\ij(\p') = \chi_\ij(\p{+}\q) 
  \simeq \chi_\ij(\p) + \q\cdot\grad_\p \chi_\ij(\p)
\end {equation}
and
\begin {equation}
  \chi_\ij(\k') = \chi_\ij(\k{-}\q)
  \simeq \chi_\ij(\k) - \q\cdot\grad_\k \chi_\ij(\k)
\end {equation}
\end {subequations}
in the collision term
(\ref{eq:collision22}).
This approximation, plus the neglect of screening,
reduces ${\cal C}^{\rm 2\lra2}/(g^4 T)$ to ${\cal A}$ times a logarithmically
divergent integral $\int dq/q$.  One or the other of these simplifications
break down outside the region $\mD \ll q \ll T$,
and so this integral more properly gives a factor of $\ln(T/\mD)$,
sometimes called a Coulomb logarithm.  For details, see
Ref.~\cite{AMY1} (or the somewhat analogous discussion of the
non-relativistic case in Ref.\ \cite{LifshitzPitaevskii}).

Now suppose that we artificially consider functions $\chi_\ij(\p)$ that are 
similar to those one finds in a leading-log calculation of transport
coefficients but which abruptly cut off above some momentum scale $p_*$
with $\mD \ll p_* \ll T$.  Specifically, consider replacing $\chi(p)$ by
the test function
\begin{equation}
  \phi^{(*)}(p) = \frac{p^\ell}{T} \, e^{-p/p_*} \,.
\label{eq:phistar}
\end{equation}
The small $\q$ approximations (\ref{eq:smallq_approx}) now break down
for $p \gtrsim p_*$ rather than $p \gtrsim T$.  Therefore, after
making the same approximations as
before, one might expect to find the same leading-log result for the
collision matrix element except for a replacement of $\ln(T/\mD)$ by
$\ln(p_*/\mD)$.
In other words,
\begin {equation}
   \Big( \phi^{(*)}_\ij \, , \, \overline{\C}\, \phi^{(*)}_\ij \Big)
  = \Big( \phi^{(*)}_\ij \, , \, \A \, \phi^{(*)}_\ij \Big)
        \Big[ \ln(p_*/\mD) + O(1) \Big] .
\label{eq:pstar_log}
\end {equation}
We will sketch a more detailed argument momentarily.
Subtracting the same matrix element of $\A \, \ln (\mu/\mD)$
from both sides, and then
taking the limit $\mD/T \to 0$ (as dictated in the definition (\ref {eq:lim1})
of $\dCbar$) with
$p_*/T$ held fixed, yields
\begin{equation}
  \frac
	{\Big( \phi^{(*)}_\ij\, , \, \dCbar(\mu/T) \, \phi^{(*)}_\ij \Big)}
	{\Big( \phi^{(*)}_\ij \, , \, \A \, \phi^{(*)}_\ij \Big)} 
	= \ln(p_*/\mu) + O(1) \, .
\label{eq:bada}
\end{equation}
Consequently,
to generate a sequence of functions demonstrating
(\ref{eq:bad}) one may simply take, for example, $p_* = e^{-m} \, T$ with
$m = 1,2,3,\cdots$.

\subsection{Details}
\label{sec:details}

To justify the estimate (\ref{eq:pstar_log}) adequately,
we must show that there
are no other contributions to the matrix element which are as large
as the small exchange momentum contribution already considered.
For simplicity of presentation, we will choose to discuss explicitly the
case where all departure from equilibrium is carried by fermions.

As a benchmark for further discussion, it is convenient first to note
the parametric size of matrix elements of $\A$.
Direct evaluation of \Eq{eq:C_LL} using $\chi(p) = \phi^{(*)}(p)$
shows that the expectation value scales like $p_*^{2\ell+1}$, so that
\begin{equation}
\Big( \phi^{(*)}_\ij , \A \, \phi^{(*)}_\ij \Big)
	\sim \frac{p^{2\ell + 1}_*}{T^3}
\label{eq:IR_value}
\end{equation}
with comparable contributions coming from gauge boson exchange
and fermion exchange
diagrams.
Now consider the contribution of $\twotwo$ processes to the same
expectation value of $\overline{\C}$.
For definiteness, start by considering
$t$-channel processes and work in the $t$-channel
representation of (\ref{eq:Q3}).
The integrand of the $\twotwo$ contribution (\ref {eq:C22}) will be
exponentially suppressed by the test function factor
$[\phi^{(*)}_\ij(\p) + \cdots - \phi^{(*)}_\ij(\k')]^2$
unless at least one of the momenta $p$, $p'$, $k$, or $k'$ are $O(p_*)$.
This is only
possible for $q-|\omega| \lsim p_*$, and either $p$ or $k$ within $O(p_*)$
of its lower integration limit.
This gives an $O(p_*^2)$
phase space suppression.
For $q\sim T$, we will have $t \sim p_* T$ and dominantly
$s \sim u \sim p_* T$ as well [because each Mandelstam variable
can be written in the form $- (P_* \pm P)^2 = \pm p_* p \, (1{-}\cos \theta)$
for some one of the external momenta $P$ with $p \sim T$].
Therefore,%
\footnote{
  We can ignore the possibility of a Bose enhancement $f_0(p_*) \sim T/p_*$
  from the external particle with $O(p_*)$ momentum that is associated
  with $\chi$, because we
  are only considering the case where the fluctuation $\chi$ is in the
  fermionic sector.
}
the contribution of $t$-channel processes from the region
$q \sim T$ to the matrix element
$\big(\phi^{(*)}, \overline{\C} \phi^{(*)}\big)$
is $O(p_*^{2\ell+2}/T^4)$.
This is smaller, by one power of $p_*/T$,
than the small-$q$ contribution which leads to
the matrix element (\ref{eq:IR_value}) for $\A$
(which also multiplies a logarithm).
Contributions from smaller exchange momenta, $q \ll T$,
will contain greater phase space suppression which can only
be compensated in those terms whose matrix elements contain
small $q$ divergences.

For $s$-channel contributions, and interference terms,
one finds the same result.
To see this explicitly,
take the matrix elements of Table \ref{table:mat_elements}
with the
substitution (\ref{eq:s2tu}) applied to the $s^2/tu$ and $u^2/st$ terms,
so that every denominator is either a power of just $s$, $t$, or $u$ (or
simply constant).  Terms with just $s$ or $s^2$ in the denominator do
not have small $s$ divergences because $t$ and $u$ become small as well.
For terms with small $t$ or $u$ divergences, we have seen that only
the small $q$ region can be non-negligible.  Consider matrix elements
that diverge as $t^{-n}$ or $u^{-n}$ with $n=1$ or 2, and focus on the
case of $t$.
For $q \sim p_*$, we have $t \sim p_*^2$, which is suppressed by a single
power of $p_*$ from the size $t \sim p_*T$ relevant for $q\sim T$. 
The contribution of these terms from $q \sim p_*$ is then enhanced by
$(T/p_*)^{n-1}$ compared to the $q \sim T$ region discussed
previously, with one power of
$p_*/T$ reflecting the greater phase space suppression on $q$.
Additionally, for the soft fermion exchange contributions to the
$qg \lra gq$ and $\bar q g \lra g \bar q$ processes,
there will be an additional Bose enhancement factor of order
$f^{\rm g}_0(p_*) \sim T/p_*$ because, for such small momentum transfers,
one of the final or initial gluons must have $O(p_*)$ momentum if one of the
initial or final quarks does.

For terms which do not involve soft fermion exchange,
the resulting contribution to the matrix element of $\overline{C}$ is therefore
$O(p_*^{2\ell-n+3}/T^{5 - n})$, which is
$(p_*/T)^{2-n}$ times the size (\ref{eq:IR_value}) of
the matrix element of $\A$.
For $qg \lra gq$ and $\bar q \lra g \bar q$ soft fermion exchange contributions,
there is an additional enhancement of $T/p_*$,
giving a result which is $(p_*/T)^{1-n}$ times (\ref{eq:IR_value}).
Since the degree of divergence $n$ is at most 2 for gluon exchange,
and 1 for fermion exchange,
this shows that even for test functions which are peaked at $p_* \ll T$, the
dominant $2 \lra 2$ matrix elements
are the underlined ones
of Table \ref{table:mat_elements} --- precisely those matrix elements
that contribute to the leading-log result.
The resulting contributions to 
$\big(\phi^{(*)}, \overline{\C} \phi^{(*)}\big)$ are indeed
dominated by $q \lesssim p_*$, as claimed earlier, and the contribution
from $q \sim p_*$
is of the same order as the result (\ref{eq:IR_value}) for $\A$.
For $q \ll p_*$, the
small $q$ divergences are softened by the cancellation of $\chi$ factors
in (\ref{eq:collision22}), which was the basis for the leading-log
approximation, reviewed earlier in discussing (\ref{eq:smallq_approx}).
As discussed previously, the contribution from $\mD \ll q \ll p_*$ then
exceeds, by a logarithm, the contribution from $q \sim p_*$.

Analyzing
$\onetwo$ processes is substantially more complicated,
but eventually leads to the result that the expectation value
$\big(\phi^{(*)}, \overline{\C}^{\onetwo} \phi^{(*)}\big)$
is $O(p_*^{2\ell+1}/T^3)$,
without any logarithmic dependence on $p_*/T$ or $\mD/T$.
Hence, $\onetwo$ contributions remain sub-leading by a logarithm
even when $p_* \ll T$.
We will only outline the analysis.
The factors of the test function $\phi^{(*)}$ in
\Eq{eq:C12} cause
the integrand to be exponentially small unless either $p$ or $k$
are $\lsim p_*$.
Symmetry of the integrand allows us to focus on $p \lsim p_*$.
The dominant contribution comes from $k \sim T$ (and so $p' \sim T$),
as can be separately
verified.  Estimating the collision operator (\ref{eq:KLPMdef}) as
${\KLPM_s}\F(\pxk) \sim g^2 \mD^2 T k^2 \nabla_\pxk^2 \F$, and noting that
$\delta E \sim \mD^2 / p + \pxk^2 / (p k^2)$, we find that the $\delta E$
term dominates, so the integral equation can be solved by expanding in
$\delta E \gg {\KLPM_s} \sim g^2\mD^2 k^2 T/\pxk^2$
with the result that $\Re \, \F \sim
{\KLPM_s}\pxk/(\delta E)^2$.  Taking $k \sim T$
and substituting into all terms in \Eq{eq:C12}, we find that
${\cal F}_q \sim p^2 g^4 T^5$,
$\gamma^q_{qg} \sim \gamma^g_{q\bar q} \sim (g^2 T)^2$, and finally obtain
$\big(\phi^{(*)}, \overline{\C}^{\onetwo} \phi^{(*)}\big)
\sim p^{2\ell+1}/T^3$.
The contribution from $k \sim p_*$ is suppressed by a further power of $p_*/T$.

\subsection {Lack of Borel summability}

Having seen that $\overline {\delta \C}$ is not relatively bounded by $\A$
[{\em c.f.} Eq.~(\ref{eq:bad})]
due to a relative logarithmic enhancement at small momentum,
we can obtain a (non-rigorous) estimate of the large order behavior
of the coefficients $Q_n$ of the inverse log expansion
by restricting the domain of $\overline {\delta \C}(\mu/T)$ to 
functions whose support lies in the momentum region $p < p_{\rm max}$
for some $p_{\rm max} \ll T$.
In other words, multiply $\overline {\delta \C}$ on both sides
by a projection onto this subspace.
Within this subspace,%
\footnote{
  In more detail, let $\phi^{(p_1)}$ represent any function roughly similar
  to (\ref{eq:phistar}) with $p_* \sim p_1$ --- that is, a function which is
  cut off for momenta large compared to $p_1 \ll T$.
  Now consider matrix elements of $\overline{\delta \C}$ of the form
  $\big(\phi^{(p_1)}, \overline{\delta \C} \, \phi^{(p_2)}\big)$.
  One can repeat the analysis of Appendix~\ref{sec:phim}
  with the only new ingredient
  being that the matrix element need not be diagonal.
  However $\overline{\delta \C}$ is diagonally-dominant:
  if $p_1 \ll p_2$, then the normalized off-diagonal matrix element
  $\big(\phi^{(p_1)}, \overline{\delta \C} \, \phi^{(p_2)}\big)/
   [\big(\phi^{(p_1)}, \phi^{(p_1)}\big)
   \big(\phi^{(p_2)}, \phi^{(p_2)}\big)]^{1/2}$
  is small compared to the normalized diagonal elements
  $\big(\phi^{(p_i)}, \overline{\delta \C} \, \phi^{(p_i)}\big)/
   \big(\phi^{(p_i)}, \phi^{(p_i)}\big)$ for $i=1,2$
  [as can be seen from the explicit form (\ref{eq:C_LL})].
  When $p_1 \sim p_2$, the momenta $p$ that dominate
  the integrals representing the matrix element will be $p \sim p_1 \sim p_2$.
  The logarithm factor in the discussion of Appendix \ref{sec:phim}
  will become $\ln(p_1/\mu) \sim \ln(p_2/\mu)$ and we have simply
  replaced this by $\ln(p/\mu)$, given which momenta $p$ dominate.
  The particular
  ordering of factors in the expression (\ref{eq:delC_approx})
  is not important at this level of precision;
  the given form is Hermitian (as it must be) and convenient for what follows.
}
\begin {equation}
    \overline {\delta\C}(\mu/T)
    \approx
    \A^{1/2} \, \bigl[ \ln (p/\mu) + K \bigr] \, \A^{1/2} \,,
\label {eq:delC_approx}
\end {equation}
for some constant $K$ which depends on just how the cutoff is implemented.
Inserting this approximation into the result (\ref {eq:Qn})
for the expansion coefficients $Q_n$, one has
\begin {equation}
    Q_{n+1}(\mu)
    \approx
    Q_{n+1}^{\rm approx}(\mu)
    \equiv
    \half \, (-1)^n \,
    \Bigl( {\cal S}_\ij ,\,
	\A^{-1/2} \> {\cal P}(p)
	\bigl[ \ln (p/\mu) + K \bigr]^n \A^{-1/2}
    {\cal S}_\ij \Bigr) \,,
\label {eq:Qn_approx}
\end {equation}
where ${\cal P}(p) = \Theta(p_{\rm max}{-}p)$
denotes the projection operator restricting the domain of $\overline{\delta\C}$.
Assuming that $p_{\rm max} < e^{-K} \, \mu$, the approximation
(\ref {eq:delC_approx}) to $\overline {\delta\C}$ is negative-definite,
implying that
the expression (\ref {eq:Qn_approx}) for $Q_{n+1}(\mu)$ is strictly positive.
Given this positivity,
we can apply the same variational approach discussed in section \ref {sec:vary}.
Specifically, using the test functions $\phi^{(*)}$,
one has the lower bound
\begin {equation}
    Q_{n+1}^{\rm approx}(\mu)
    \ge
    \frac{
    \half \,
    \Big( {\cal S}_\ij , \, \phi^{(*)}_\ij \Big) 
    \Big( \phi^{(*)}_\ij , \, {\cal S}_\ij \Big) 
    }
    {
    \Big( \phi^{(*)}_\ij , \, \A^{1/2} \,
	\bigl[ -\ln (p/\mu) - K \bigr]^{-n} \A^{1/2}
    \, \phi^{(*)}_\ij \Big) 
    } \,.
\end {equation}
Consider now, for simplicity, the case of
$\ell=1$ with all departure from equilibrium carried by
fermions, as is relevant for flavor diffusion.
An easy estimate shows that
\begin{equation}
    \Big( {\cal S}_i , \, \phi^{(*)}_i \Big) \sim p_*^4 \, , 
\end{equation}
while from Eq.~(\ref{eq:IR_value}) one has
\begin {equation}
    \Big( \phi^{(*)}_i , \,
	\A^{1/2} \bigl[ -\ln (p/\mu) - K \bigr]^{-n} \A^{1/2}
    \, \phi^{(*)}_i \Big) 
    \sim
    p_*^3 \> \bigl[ -\ln (p_*/\mu) - K_1 \, \bigr]^{-n} \,,
\end {equation}
for some new constant $K_1$.
Hence
\begin{equation}
    Q_{n+1}^{\rm approx}(\mu)
    \ge
    K_2 \> p_*^5 \, \bigl[ -\ln (p_*/\mu) - K_1 \bigr]^n \, ,
\end{equation}
for some positive constant $K_2$.
(The $O(1)$ constants $K_1$ and $K_2$ do not depend on $p_*$ or $n$.)
Since this is a lower bound, 
we may maximize over all values of $p_*$
(subject to $p_* < e^{-K_1} \, \mu$),
which gives $p_*|_{\rm opt} = e^{-n/5} \, e^{-K_1} \, \mu$ and
results in the estimate
\begin{equation}
  Q_{n+1}^{\rm approx}(\mu) \ge
  K_2 \, (e^{-K_1} \mu)^5 \, e^{-n} \left( \frac{n}{5} \right)^n \, .
\label {eq:Qbound}
\end{equation}
The case of shear viscosity,
where $\ell = 2$ and bosons are also out of equilibrium, is similar.
The non-alternating factorial growth of (\ref {eq:Qbound}) with $n$
immediately implies that the inverse log expansion (\ref {eq:Qseries})
has zero radius of convergence,
and moreover that its Borel transform has a singularity on the positive
real axis (at 5).
Although the estimate (\ref {eq:Qbound}) is based on the
approximation (\ref {eq:delC_approx}),
we do not see how inclusion of the non-logarithmically enhanced parts of
$\overline {\delta\C}$ can affect the $n!$ growth of the above estimate.
Assuming that this estimate does give the correct large order behavior
of the inverse log expansion,
it is worth noting that the resulting renormalon ambiguity associated with
the singularity of the Borel transform
would only be of order $(\mD/T)^5 \sim g^5$.

\end{document}